\documentclass[12pt,a4paper]{article}
\pdfoutput=1
\usepackage[utf8]{inputenc}
\usepackage{amsmath}
\usepackage{amsfonts}
\usepackage{amssymb}
\usepackage[left=2.5cm,right=2.5cm,top=2cm,bottom=2.5cm]{geometry}
\usepackage{slashed}
\usepackage{graphicx}
\usepackage{caption}
\usepackage{slashed}
\usepackage{float}
\usepackage{subcaption}
\usepackage{jheppub}

\hypersetup{
	bookmarks=true,         % show bookmarks bar?
	unicode=true,          % non-Latin characters in Acrobat's bookmarks
	pdftoolbar=true,        % show Acrobat's toolbar?
	pdfmenubar=true,        % show Acrobat's menu?
	pdffitwindow=false,     % window fit to page when opened
	pdfstartview={FitH},    % fits the width of the page to the window
	pdftitle={ForbiddenFI},    % title
	pdfauthor={Author},     % author
	pdfsubject={Subject},   % subject of the document
	pdfcreator={LD, DK, AH, LR},   % creator of the document
	pdfproducer={Producer}, % producer of the document
	pdfkeywords={dark matter} {freeze-in} {forbidden}, % list of keywords
	pdfnewwindow=true,      % links in new PDF window
	colorlinks=true,       % false: boxed links; true: colored links
	linkcolor=black,          % color of internal links (change box color with linkbordercolor)
	citecolor=black,        % color of links to bibliography
	filecolor=black,      % color of file links
	urlcolor=black           % color of external links
}

\usepackage{color}
\newenvironment{bl}[1]{{\color{blue} #1}}

\newcommand{\mST}{m_{S,T}{}}
\newcommand{\mSV}{m_{S}{}}
\newcommand{\ysx}{y_{\chi}}

\newcommand{\omh}{\Omega h^{2}}
\newcommand{\Ycr}{ Y_{ \rm DM, 0 }{} }
\newcommand{\Yc}{ Y_{\rm DM}{} }
\newcommand{\mc}{ m_{\chi} }
\newcommand{\TR}{ T_{\rm R} }
\newcommand{\zR}{ z_{\rm R} }
\newcommand{\xR}{ x_{\rm R} }

\newcommand{\ie}{{\em i.e.}}    %lr \newcommand{\ie}{{\em i.e.}~}
\newcommand{\eg}{{\em e.g.}}    %lr \newcommand{\eg}{{\em e.g.}~}
\newcommand{\GeV}{{\rm GeV}}
\newcommand{\TeV}{{\rm TeV}}
\newcommand{\MeV}{{\rm MeV}}

\newcommand{\vev}[1]{\langle #1 \rangle}
\newcommand{\lrb}[1]{\left( #1 \right)}
\newcommand{\lrsb}[1]{\left[ #1 \right]}

\usepackage{ifthen}
\usepackage{tikz}

\newcommand{\eqs}[1]{\foreach\i[count=\NumArgs] in {#1}{}%
\ifthenelse{\equal{\NumArgs}{1}}{eq.~(\ref{#1})}%
{\ifthenelse{\equal{\NumArgs}{2}}%
{eqs.~\foreach\i[count=\q]in{#1}{\ifthenelse{\equal{\q}{\NumArgs}}{and (\ref{\i})}{(\ref{\i})~}}}%
{eqs.~\foreach\i[count=\q]in{#1}{\ifthenelse{\equal{\q}{\NumArgs}}{and (\ref{\i})}{(\ref{\i}),~}}}}}

\newcommand{\Eqs}[1]{\foreach\i[count=\NumArgs] in {#1}{}%
\ifthenelse{\equal{\NumArgs}{1}}{eq.~(\ref{#1})}%
{\ifthenelse{\equal{\NumArgs}{2}}%
{Eqs.~\foreach\i[count=\q]in{#1}{\ifthenelse{\equal{\q}{\NumArgs}}{and (\ref{\i})}{(\ref{\i})~}}}%
{Eqs.~\foreach\i[count=\q]in{#1}{\ifthenelse{\equal{\q}{\NumArgs}}{and (\ref{\i})}{(\ref{\i}),~}}}}}

\newcommand{\refs}[1]{\foreach\i[count=\NumArgs] in {#1}{}%
\ifthenelse{\equal{\NumArgs}{1}}{(\ref{#1})}%
{\ifthenelse{\equal{\NumArgs}{2}}%
{\foreach\i[count=\q]in{#1}{\ifthenelse{\equal{\q}{\NumArgs}}{and (\ref{\i})}{(\ref{\i})~}}}%
{\foreach\i[count=\q]in{#1}{\ifthenelse{\equal{\q}{\NumArgs}}{and (\ref{\i})}{(\ref{\i}),~}}}}}
\title{Forbidden frozen-in dark matter}

\author[a]{L. Darmé,}
\author[a]{A. Hryczuk,}
\author[a]{D. Karamitros,}
\author[b,a]{L. Roszkowski}

\affiliation[a]{ National Centre for Nuclear Research, ul. Pasteura 7, 02-093 Warsaw, Poland}
	
\affiliation[b]{Astrocent, Nicolaus Copernicus Astronomical Center Polish Academy of Sciences,  Bartycka 18, 00-716 Warsaw}

\emailAdd{luc.darme@ncbj.gov.pl}
\emailAdd{andrzej.hryczuk@ncbj.gov.pl}
\emailAdd{dimitrios.karamitros@ncbj.gov.pl}
\emailAdd{leszek.roszkowski@ncbj.gov.pl}

\abstract{ We examine and point out the importance of a regime of dark
  matter production through the freeze-in mechanism that results from a
  large thermal correction to a decaying mediator particle mass from
  hot plasma in the early Universe.  We show that mediator decays to
  dark matter that are kinematically forbidden at the usually
  considered ranges of low temperatures can be generically present at
  higher temperatures and actually dominate the overall dark matter production,
  thus leading to very distinct solutions from the standard case.  We
  illustrate these features by considering a dark Higgs portal model
  where dark matter is produced via decays of a scalar field with a
  large thermal mass.  We identify the resulting ranges of parameters
  that are consistent with the correct dark matter relic abundance and
  further apply current and expected future collider, cosmological,
  and astrophysical limits.
}

\begin{document}
\maketitle
\flushbottom

%%%%%%%%%%%%%%%%%%%%%%%%%%%%%%%%%%%%%%%%%%%%%%%%%%%%%%%%%%
\section{Introduction}
\label{sec:intro}

Attempts to explain the presence and abundance of dark matter (DM) in the
Universe often involve making various assumptions about the history of
the very early Universe. The simplest and most natural one is to
assume that, at high enough temperatures a DM particle is in
thermal equilibrium with the plasma of Standard Model (SM) particles,
which ensures that its density is given by Maxwell-Boltzman
statistics. At some point in the expansion and cooling down of the
Universe, DM undergoes a well-known freeze-out mechanism, which
determines its subsequent population in the Universe. The freeze-out
mechanism has been particularly popular because it requires a minimum
amount of rather natural assumptions and, for reasonable values of
parameters of specific particle candidates in the class of
weakly-interacting massive particles (WIMPs), it is often able to
produce the observed abundance of DM in the Universe. Furthermore, it
does so in a manner that is insensitive to the condition of the Universe 
after inflation, thus effectively separating the high temperature regime 
from the one responsible for dark matter production.

However, it has long been known that, in addition to freeze-out, some
other DM production mechanisms exist and can in fact play a dominant
role in achieving the observed relic density. One particularly
well-motivated example involves sub-eV axions that, due to their tiny
interactions, are mainly produced not thermally but via the well-known
misalignment mechanism; for recent reviews see,
\eg,~\cite{Baer:2014eja,Marsh:2015xka}. This mechanism was later
extended to the case of ultra-light vector boson
in~\cite{Nelson:2011sf,Arias:2012az}. 

Furthermore, extremely weakly interacting massive particles (usually
referred to as E-WIMPs or super-WIMPs) are often predicted by many
well-motivated extensions of the SM, for instance a gravitino in
scenarios based on local supersymmetry (SUSY) or an axino in SUSY
models of axions; see \eg,~\cite{Baer:2014eja} for a recent review. If stable, they are potential candidates for dark matter in the
Universe. However, due to their exceedingly feeble interactions, their
population after inflation is negligible -- assuming that their
decoupling temperature is higher than the reheating temperature
$T_{\rm R}$ -- since they never reach thermal equilibrium with the SM
plasma, and the freeze-out mechanism is ineffective. Instead, they can
be generated through so-called freeze-in~\cite{Hall:2009bx} from
scatterings and decays of some other particles.

A key feature of such ``frozen-in'' dark matter scenarios is that,
while all SM particles remain in thermal equilibrium since the
Universe reheats after inflation, the DM particle $\chi$ is absent in
the early Universe and never reaches equilibrium with the SM
plasma. Its
production is mediated by some particles that typically remain in
equilibrium with the plasma. Once the temperature drops below the mediator mass, DM
production essentially stops and its relic density freezes-in.  

In freeze-in scenarios, specific features and the final relic abundance of DM often
depend on the details of a specific beyond-the-SM (BSM) model. In
models with either the gravitino or axino as DM, their freeze-in production is
typically dominated by non-renormalizable interactions at high
temperatures in the case of scattering or at low ones in the case of
decays~\cite{Ellis:1984eq,Covi:2001nw}. On the other hand, in models where DM production
involves for instance a light mediator, the low-temperature production dominates
over the high-temperature one, thus separating again the physics of
inflation from the one of dark matter~\cite{McDonald:2001vt,Hall:2009bx,Blennow:2013jba}.

In this article, we will consider a previously neglected case that
some mediator field $S$ -- that could be a scalar, vector boson or a fermion
-- is not only in equilibrium with the thermal bath, but also develops
a substantial thermal mass. That is, at sufficiently high temperatures
the mass $\mST$ of the mediator deviates significantly from its
``vacuum" one $\mSV$, \ie, the mass is dominated by thermal
effects. Such an effect has recently been studied for instance while
considering thermal photon decays~\cite{Dvorkin:2019zdi}. The
population of DM particles $\chi$ is assumed to be initially absent
when the Universe reheats after inflation, but is generated by the
decays of $S$. If at high enough temperatures the thermal mass of the
mediator becomes sufficiently large, the possibility opens up that,
when $\mST > 2m_{\chi}$ the decay $S \to \bar{\chi} \chi$ becomes
allowed, while at $T=0$ it was kinematically forbidden. As we will
show, this opens up a new regime for DM production which we will call
\textit{``forbidden frozen-in dark matter''}.

This kind of effect we believe was first identified for gravitino~\cite{Rychkov:2007uq}
and subsequently axino production~\cite{Strumia:2010aa}. 
The production rate of singlet fermions from the decay of scalar fields in a  plasma including thermal corrections was calculated in~\cite{Drewes:2015eoa} and applied to the case of right-handed neutrino DM.
More recently it was also described in a
more generic context in~\cite{Baker:2017zwx,Bian:2018mkl}.

In this paper, we take a closer look at the ``forbidden freeze-in''
regime and identify its main phenomenological features.  We further
show that the equilibrium assumption of the mediator can be relaxed as
long as $S$ obtains a sizeable thermal correction to its mass, \eg,
when it is chemically decoupled from the SM plasma, but remains in
kinetic equilibrium with itself via self-scatterings. Interestingly,
albeit perhaps as expected, the ensuing phenomenology is found to
depend strongly on the dimension of the operator controlling the
mediator decay into DM pair. For dimension-four operators, the
production is dominated at low-temperature regime and peaks at $\mST \sim
2m_{\chi}$. Additionally, a striking feature is that, in this regime
the relic abundance is ultimately almost insensitive of the DM mass,
while the coupling responsible for DM production typically takes
significantly larger values than in the standard freeze-in case. For
mediator decays through higher-dimensional operators, on the other
hand, the production is dominant at high temperatures and
therefore depends on the reheating temperature.
Furthermore, we argue that, since the forbidden freeze-in regime is a
generic property, it might be worth exploring it in models of the
freeze-in mechanism of DM production,
\eg~\cite{Aoki:2015nza,Ayazi:2015jij,Shakya:2015xnx,Tsao:2017vtn,Dedes:2017shn,Bae:2017dpt,Duch:2017khv,
  Biswas:2018aib,Bhattacharyya:2018evo,Goudelis:2018xqi,Belanger:2018sti,Abdallah:2019svm}.

As a specific realisation of the case presented above, we examine an
explicit Higgs portal scenario, where the dark Higgs boson, kept in
equilibrium with the SM fields through a quartic mixing term with the SM
Higgs, can decay to a light (GeV-scale) Dirac fermion dark matter at a
strongly suppressed rate. The thermal mass is predominantly generated by
the dark Higgs self-coupling, enabling it to easily reach a thermal-mass
dominated regime. Since Higgs portal scenarios are typically
constrained by a variety of limits, we briefly review them and apply
them to the considered model provide in order to identify new regions
that are allowed by forbidden freeze-in. We will assume the dark Higgs boson is originally in equilibrium with the SM thermal bath. If the quartic mixing is low enough, it may also be produced through freeze-in,  see \eg,~\cite{Heeba:2018wtf} for more details on this setup. %% ADDED
This scenario is similar to case III of~\cite{Baker:2017zwx}, which
closely resembles our model as the particle content is similar. The
main difference, however, is the vacuum expectation value (VEV)
structure. In our model the portal particle has a zero VEV
($\vev{S}=0$) throughout the early Universe, and develops one only
through its mixing with the Higgs boson after electroweak phase
transition. This allows us to isolate the pure forbidden freeze-in
regime without the impact of the SM Higgs VEV, thus simplifying the
analysis and exploring the forbidden freeze-in independently. We additionally explore a different mass region than~\cite{Baker:2017zwx}, which leads to a distinct phenomenology for the portal particle.

The paper has the following structure. In section
Sec.~\ref{sec:forbfreeze-in} we briefly review the calculation of
thermal mass of a scalar boson and proceed to describe in detail the
mechanism of ``forbidden freeze-in'' through thermal mass effects. In
Sec~\ref{sec:DarkHiggs} we consider as an example an explicit
Higgs-portal model in which the scenario can natually be realised, and
briefly examine various criteria to ensure its consistent
implemention. We then proceed to a full numerical study of the
predicted relic density and describe various aspects of our scans and
results, as well as the effect of applying relevant astrophysical and
collider constraints.

%%%%%%%%%%%%%%%%%%%%%%%%%%%%%%%%%%%%%%%%%%%%%%%%%%%%%%%
\section{Freeze-in with a thermally induced mass}\label{sec:forbfreeze-in}

\subsection{Thermal mass in the early Universe }\label{subsec:thermal_mass}
\setcounter{equation}{0}

As mentioned above, in this article we study the freeze-in production
of DM via some mediator decays that are energetically allowed solely
in a thermal bath.  We expect this to occur in general, since
frozen-in DM is usually assumed to be produced by particle species
which are in thermal equilibrium with the SM plasma, and which should
therefore develop a thermal mass
correction~\cite{Das:1997gg,Kapusta:2006pm,Bellac:2011kqa} in the
early Universe, similarly to the SM particles~\cite{Fodor:1994bs}.
Moreover, it is this effective mass that allows ``forbidden" decays to
occur, as is the case for instance for plasmons (thermally-dressed
photons in a medium) that can decay to neutrinos~\cite{Braaten:1993jw}.

Generally, at high temperatures applicable to the early Universe
the thermal mass of a particle is proportional to the temperature.  
As this effect will be critical in realizing our forbidden freeze-in scenario, 
below we briefly review the case of a scalar mediator field $S$.

%
%%%%%%%%%%%%%%%%%%%%%%%%%%%%%%%%%%%%%%%%%%%%%%%%%%%%%%%
\begin{figure}[h!]
	\centering
	\includegraphics[width=0.4\textwidth]{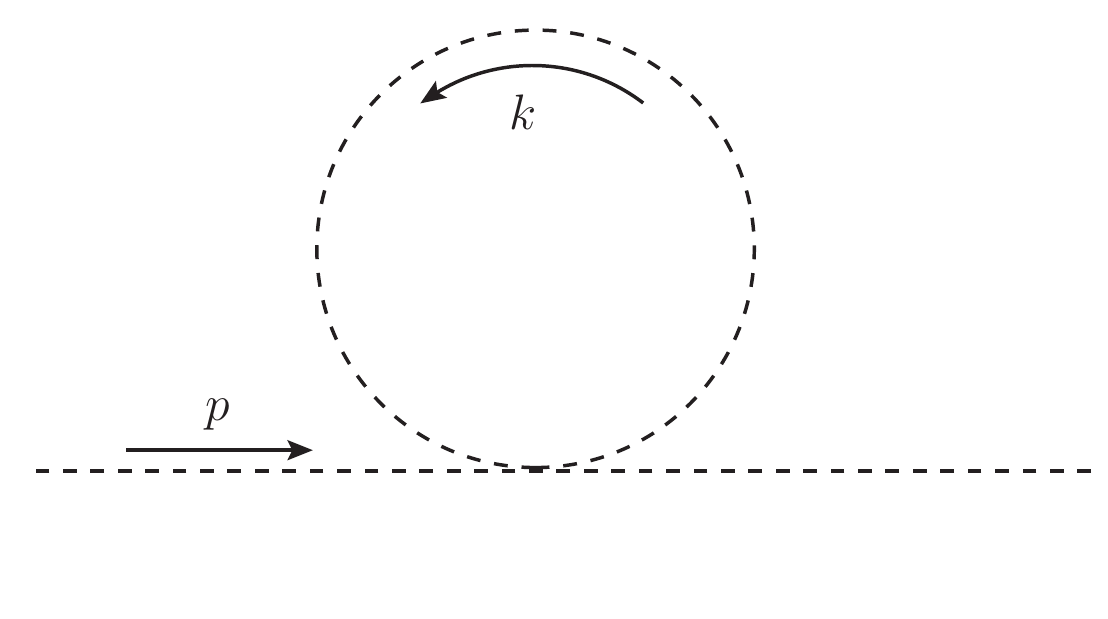}
	%\vspace{-0.8cm}
		\vspace{-0.3cm}
	\caption{One-loop self-energy for the scalar induced by its self-interaction.}  
	\label{fig:self_energy}
\end{figure}
%%%%%%%%%%%%%%%%%%%%%%%%%%%%%%%%%%%%%%%%%%%%%%%%%%%%%%%%%%%

In general, a scalar field features a self-interaction term, which implies
that it does not need to interact very strongly with the rest of the
plasma in order to develop a sizeable thermal mass. In the following we assume a
self interaction term for $S$ of the form
\begin{equation}
\mathcal{L}_{S}=-\dfrac{\lambda_{S}}{4!} \; S^{4} \, .
\label{eq:S^4}
\end{equation}
The self-energy diagram, shown in Fig.~\ref{fig:self_energy}, can then
be readily evaluated at a finite temperature $T$, leading to the
self-energy term
\begin{equation*}
\Pi_{S}=\dfrac{\lambda_{S}}{2\beta} \sum_{n=-\infty}^{\infty}\int \dfrac{ d^{3} \vec{k} }{ (2\pi)^{3} } \dfrac{1}{\omega_{n}^{2} - \omega_{k}^{2}},
\end{equation*}
where $\Pi_{S}$ corresponds to the corrected mass of $S$, \ie,
$\mST^2=\mSV^{2}+\Pi_{S}$, and we have denoted $\beta=T^{-1}$,
$\omega_{n}=2n\pi \beta^{-1}$, and $\omega_{k}^{2}=\vec{k}^{2} +
\mSV^{2}$.

The sum over $n$ is evaluated by a standard
procedure:\footnote{Details can be found in the literature, \eg,
\cite{Das:1997gg,Kapusta:2006pm,Bellac:2011kqa}. } by transforming
it to an integral over a complex quantity $\omega$ while introducing a
function which has poles corresponding to $\omega_{n}$ and unit
residue. One obtains
\begin{eqnarray*}
\Pi_{S}=i\dfrac{\lambda}{2} \int \dfrac{d^{4}k}{(2\pi)^{4}} \dfrac{1}{k^{2}-\mSV^{2} + i\epsilon} + 
\dfrac{\lambda}{2} \int \dfrac{ d^{3} \vec{k} }{ (2\pi)^{3} } \dfrac{f_{B} (\omega_{k})}{\omega_{k}} \, ,
\end{eqnarray*}
where we identify the first term as the $T=0$ one-loop correction to $m_{S}$, and  the second one (denoted $\Pi_{S}^{(T)}$ henceforth) 
as the correction due to the finite temperature
of the medium with 
$f_{B} ~\equiv~\left(e^{\omega_{k} \beta}-1\right)^{-1}$ 
the Bose-Einstein phase-space distribution. The appearance of the
phase-space distribution function regulates this otherwise
quadratically divergent integral since it introduces a natural
``cut-off" energy proportional to the temperature.  The final result
scales quadratically with temperature: $\Pi_{S}^{(T)} \sim
T^{2}$. In the high temperature limit, we can therefore neglect the
$\mSV$ contribution to $\omega_{k}$ and arrive at
\begin{equation}
\Pi_{S}^{(T)} = \dfrac{\lambda}{24}T^{2} \, .
\label{eq:Pi_{S}^{T}}
\end{equation}
In this limit, since the vacuum one-loop contribution is expected to be small
compared to the tree-level one, we can neglect all $T=0$
contributions and obtain an estimated form of the mass of $S$,
\begin{equation}
\mST^{2} \approx \Pi_{S}^{(T)}= \dfrac{\lambda_{S}}{24}T^{2} \, .
\label{eq:mST}
\end{equation}

It is well known, though, that naive perturbation theory does not
work well when finite temperature effects are included (for examples
see~\cite{Das:1997gg,Kapusta:2006pm,Bellac:2011kqa}). This can be seen
by calculating the thermal correction using $m_{S}^{2} \to
\dfrac{\lambda}{24}T^{2}$, \ie, by re-summing the so-called ``daisy"
diagrams, where one would expect to get a correction of order at least
$\mathcal{O}(\lambda^{2})$. However, this is not the case in finite
temperature calculations, since such diagrams induce correction
$\mathcal{O}(\lambda^{3/2})$, which may be important especially for
larger values of the self-interaction coupling. We have explicitly checked
that for $\lambda \lesssim 1$ this re-summation leads to at
most  a $20 \%$ variation in the thermal mass. We will thus use the
approximate result~\eqs{eq:mST} throughout this paper.

%%%%%%%%%%%%%%%%%%%%%%%%%%%%%%%%%%%%%%%%%%%%%%%%%%%%%%%
%\subsection{Forbidden freeze-in}\label{subsec:std_freeze-in}

\subsection{Freeze-in and mediator decay}\label{subsec:std_freeze-in}
We are interested in estimating the final relic density of a DM
particle $\chi$ interacting extremely feebly with the Standard Model
particles. The key assumption is that $\chi$ was never in thermal
contact with the SM sector during the thermal history of the Universe,
nor was it ever produced through some other means in the
post-inflationary period, \eg, during reheating. Our assumed dominant
dark matter production mechanism will be a suppressed decay of a bath
particle $S$ into a dark matter pair. More precisely, following the
standard lore, we will assume the presence of a strongly suppressed
decay channel
\begin{align}
\label{eq:Meddecay}
	S \rightarrow \bar{\chi}\chi   \ ,
\end{align}
with a small decay rate $\Gamma_\chi$ (\ie, such as to unable one to
overproduce or thermalise the $\chi$s). Assuming a boson mediator and
neglecting Pauli blocking/Bose-Einstein enhancement factors, the
Boltzmann equation governing the density of dark matter particle in an
expanding universe is then (see, \eg,~\cite{Belanger:2018mqt} for a
complete recent treatment) given by
\begin{align}
	\dot{n}_\chi + 3 H n_\chi = \int d \Pi_S d \Pi_\chi d \Pi_{\bar{\chi}} \times \frac{1}{e^{E_S/T}-1} \times (2 \pi)^4 \delta^4 (P_S - P_\chi -P_{\bar{\chi}}) \displaystyle \sum_{\rm idof's}	|\mathcal{M}|^2 \, , 
\end{align}
where $\mathcal{M}$ is the amplitude (summed over all internal degrees
of freedom ``idof'') for the decay process~\eqref{eq:Meddecay}, and
the integration is over the standard phase space factors $d \Pi_{S}
\equiv \displaystyle \frac{d^3 p_{S}}{(2\pi)^3 2 E_{S}}$, and
similarly for $d \Pi_{\chi}$ and $d \Pi_{\bar{\chi}} $. Without loss
of generality regarding the operator generating the decay $S
\rightarrow \bar{\chi}\chi$, we can rewrite the squared amplitude from
the decay rate $\Gamma_\chi$ as
\begin{align}
\displaystyle \sum_{\rm idof's}| \mathcal{M}|^2   = (2 J_S +1 ) \frac{8 \pi m_S^3}{\sqrt{\lambda(m_S^2,m_\chi^2,m_\chi^2)}} \Gamma_\chi \, ,
\end{align}
where $\lambda$ is the usual K\"all\'en/triangle function and $J_S$ is
the spin of the mediator. It can then be shown (see, \eg,
\cite{Belanger:2018mqt}) that under some general assumptions (\ie, a
negligible initial number of DM particles, entropy conservation, and
Maxwell-Boltzmann distributions for the plasma), the evolution of the
DM yield ($\Yc=\frac{n_{\chi}+n_{\bar{\chi}}}{s}$) is given by
\begin{equation}
-HsT \, \delta_{h}^{-1} \frac{d\Yc}{dT}=\dfrac{   (2 J_S +1 ) \Gamma_\chi}{ \pi^2} \, K_{1}( m_{S}/T) \; m_{S}^{2} \; T \, .
\label{eq:dYdT_std}
\end{equation}
with
\begin{align}
	s~\equiv~&\dfrac{2 \pi^{2}}{45} h(T) \, T^{3}\, , \\
	\delta_{h}~\equiv~& 1+\dfrac{1}{3}\dfrac{d\, \log(h)}{d\, \log(T)},\\
	H~=~&\sqrt{\dfrac{4 \pi^{3}}{45 \, m_{P}^{2}} g(T) } ~T^{2} \, ,
\end{align}
where $h$ ($g$) are the relativistic degrees of freedom associated
with the entropy (energy) density,\footnote{To obtain our numerical
results we use the ones provided in~\cite{Drees:2015exa}. } and $K_{1}(x)$ 
the modified Bessel function of the first kind. Defining $x\equiv \dfrac{m_{S}}{T}$ 
and focusing for simplicity on $J_S=0$, the evolution of the yield becomes
\begin{equation}
\dfrac{d\Yc}{dx}=\lrb{ \dfrac{\Gamma_\chi (m_S,m_\chi) }{ 5.93 \times 10^{-19} \, \GeV} } \lrb{ \dfrac{1 \  \GeV}{m_S} }^2 
\dfrac{K_{1}(x) x^{3}}{\sqrt{g}h} \, \delta_{h} \, .
\label{eq:dYdx_std}
\end{equation}

An important comment at this point is that, while in the standard
freeze-in case $\Gamma_\chi $ can be considered to be a number which
factors out of the $x$ dependence, this is not the case for forbidden 
freeze-in where the presence of a thermal mass $m_S(T)$ needs to be accounted for. Let us first review in the rest of this section the standard freeze-in case where the thermal dependence of the mass can be neglected.

% \subsubsection*{Reviewing the standard freeze-in scenario}

 Assuming
that $\mSV > 2 m_\chi$ and slowly varying relativistic degrees of
freedom (which is the case for $T \gtrsim 1 ~\GeV$), we can calculate
the yield today ($\Ycr$), by integrating from the reheating
temperature ($\TR \gg \mSV$, $x \to 0$) down until today ($T_{0} \ll
\mSV$, $x \to \infty$).\footnote{If DM is produced mainly at
temperatures at which the assumptions are violated, $\Ycr$ can
obtained numerically from eq.~\eqref{eq:dYdx_std}.} We then obtain
the relic abundance in the form
\begin{equation}
\omh\approx  2.8 \times 10^{8} \, \dfrac{m_{\chi}}{\GeV} \Ycr \approx \lrb{ \dfrac{\Gamma_\chi (m_S,m_\chi) }{ 4.5 \times 10^{-28} \, \GeV} } \lrb{ \dfrac{1 \  \GeV}{m_S} }^2 
m_{\chi} \lrb{ \dfrac{1}{ \sqrt{g} \; h }  } \! \Big{|}_{x=\vev{x} } \, ,
\label{eq:relic_def}
\end{equation}
where we evaluate $g$ and $h$ at the ``mean" value of $x$ during the 
DM production.\footnote{This value is defined as
\begin{equation*}
\vev{x}\equiv \dfrac{\int_{0}^{\infty} dx \, x^{3} K_{1}(x) \times x }{\int_{0}^{\infty} dx \, x^{3} K_{1}(x)}  \approx 3.4 \, .
\end{equation*}
}
%%%%%%%%%%%%%%%%%%%%%%%%%%%%%%%%%%%%%%%%%%%%%%%%%%%%%%%
\begin{figure}[t]
	\centering
	\begin{subfigure}[b]{0.49\textwidth}
		\centering
		\includegraphics[width=1\textwidth,height=0.85\textwidth]{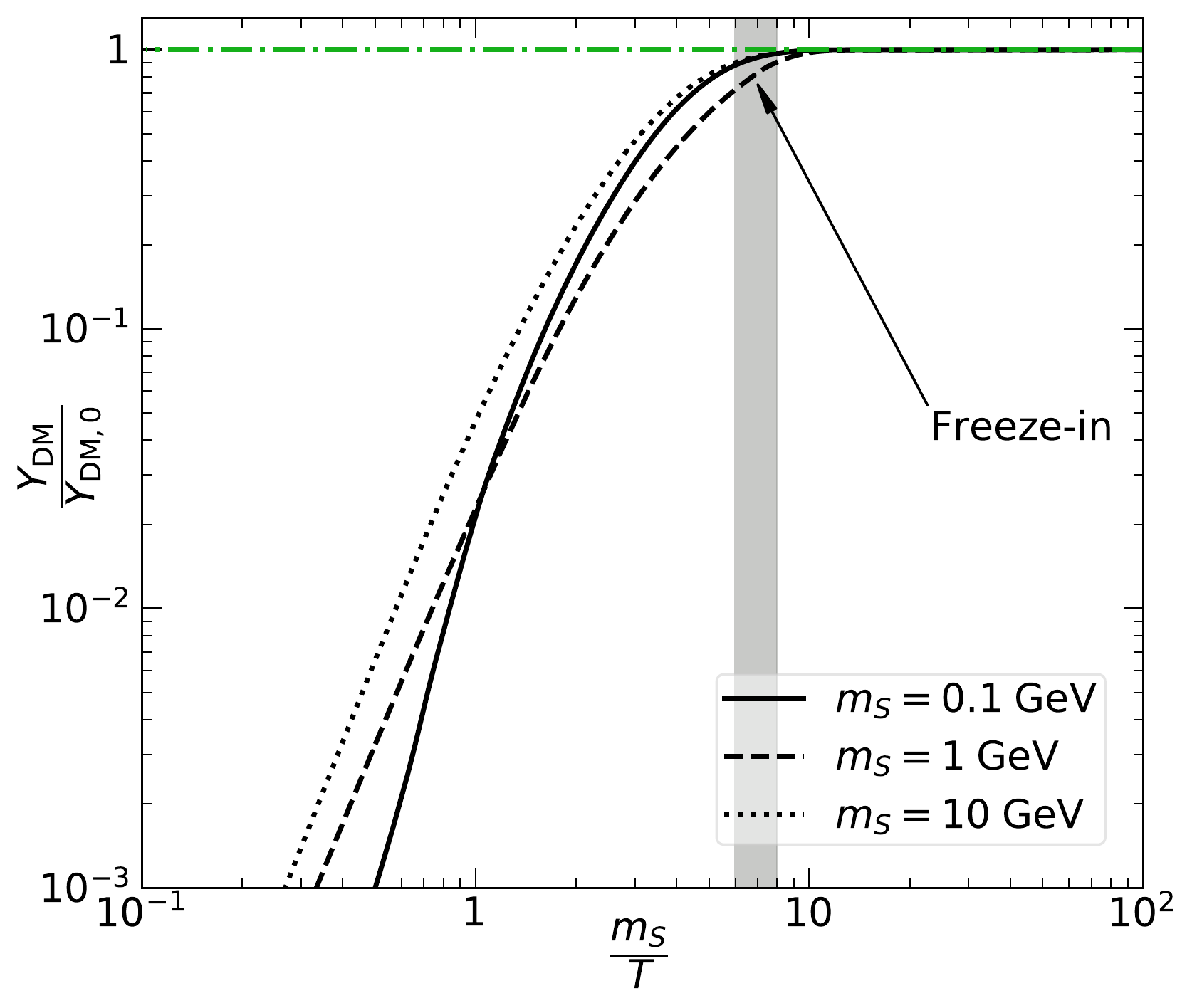}
		\caption{}
		\label{fig:Y_ev_std}
%		\vspace*{0.7mm}
	\end{subfigure}
%		\hfill
	\begin{subfigure}[b]{0.49\textwidth}
		\centering
		\includegraphics[width=1\textwidth,height=0.85\textwidth]{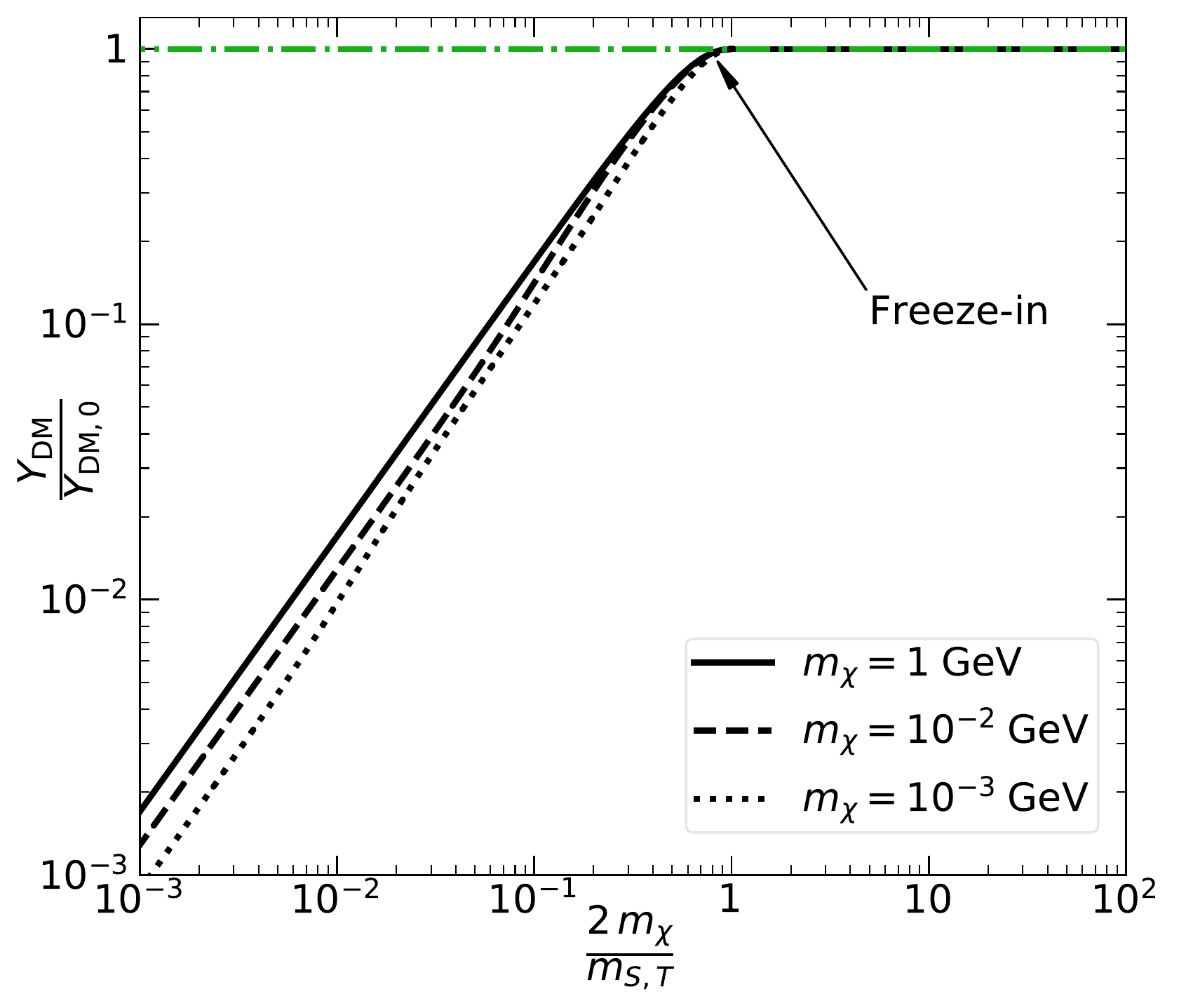}
		\caption{}
		\label{fig:Y_ev_frb}
	\end{subfigure}
	\caption{Typical evolution of $\Yc/\Ycr$  during the
		production of DM.  {\bf (a)} The evolution as a function of $x
		\equiv m_{S}/T $ for $m_{S}=0.1, \, 1, \,10 \; \GeV$, and
		$m_{\chi}=0$ (\ie, {\em standard freeze-in}). The gray area
		denotes where the freeze-in occurs.  {\bf (b)} The evolution as a
		function of $z\equiv 2\mc/\mST$ for $\mc=10^{-3}, \,
		10^{-2}, \,1 \; \GeV$ , and $\mSV=0$ (\ie,  forbidden
		freeze-in).  In both figures, green lines show 	$\Yc/\Ycr=1$.	}
	\label{fig:Y_evolution}
\end{figure}
In Fig.~\ref{fig:Y_ev_std} we show $\Yc/\Ycr$ as a function of $x$ for 
various values of $m_{S}$, where we see that the production of DM 
essentially stops at the freeze-in temperature $T_{\rm FI} \sim
\frac{m_S}{7}$, as can be seen from the figure.
That is, since typically $S$ decouples at temperature $T_{\rm FO}
\approx 20 \, m_{S}$ (\ie, freeze-out), the calculation holds.
However, if $S$ decouples earlier than expected, the relic abundance
of $\chi$ can be considerably smaller (if $S$ decays rapidly to SM
particles) or larger (if $S$ decays predominantly to DM particles). In
both cases the coupled system of Boltzmann equations describing the
evolution of both $S$ and $\chi$ has to be solved.

A different behavior is expected, however, when DM particles are
produced via non-renormalizable operators, since the corresponding
production rate increases with the
temperature~\cite{Hall:2009bx,Elahi:2014fsa}. As an example, consider
DM production via a $2 \to 2$ process which occurs due to a
dimension-$d$ operator. At high temperatures, all masses should be
irrelevant, so the matrix element squared for the process can be
written as function of the center-of-mass energy $\sqrt{\hat{s}}$ as
\begin{eqnarray*}
	|\mathcal{M}|^{2} \approx \gamma_d \lrb{\dfrac{\sqrt{\hat{s}}}{\Lambda}}^{2n} \, , 
\end{eqnarray*}
with $n=d-4$. The corresponding Boltzmann equation is
\begin{equation*}
	\dfrac{d\Yc}{dx} \approx \dfrac{1}{512 \pi^{5}} \ \dfrac{\delta_{h}}{H \, s \, x }  \dfrac{m_S}{x}\int_{0}^{\infty} d\hat{s} \;  \dfrac{\hat{s}^{n+1/2}}{\Lambda^{2n}} \;  K_1\lrb{\dfrac{\sqrt{\hat{s}} }{T} x } \, ,   
\end{equation*}
which (assuming constant $g$ and $h$) can be integrated from $\xR=m_{S}/\TR$ to today ($x_{0}$). The result is
\begin{equation}
	\Ycr \approx \dfrac{\xR^{1-2n} - x_{0}^{1-2n}  }{ 2n-1} \lrb{ \dfrac{ 4^{n} \, n! \, (n+1)! \, \gamma_{d}}{2.34 \times 10^{-15}} }   
	\lrb{\dfrac{m_{S}}{\Lambda}}^{2n}  \lrb{\dfrac{1 \ \GeV}{m_{S}} } \lrb{ \dfrac{1}{ \sqrt{g} \; h }  } \! \Big{|}_{x \sim \xR }\, ,
	\label{eq:Y_std_UV}
\end{equation}
where it is apparent that the high-temperature contributions dominate
for $n>0$ (\ie, $d>4$). In the case of $d \leq 4$, we expect DM
production to be dominated at low temperatures (around $m_S$, as denoted
previously). Thus, these features should be treated in a case-by-case way,
since the  masses of the particles play an important role, and so
the actual structure of the matrix element is needed.

%%%%%%%%%%%%%%%%%%%%%%%%%%%%%%%%%%%%%%%%%%%%%%%%%%%%%%%%%%%%%%%%%%%%%%%%%%%%%%

\subsection{Large thermal mass and forbidden freeze-in}
Let us now turn to the case with a large thermal mass. In order to
determine its effect on the freeze-in mechanism we shall assume for
concretness that the scalar mediator mass takes the form\footnote{In
  the case where $S$ gets its thermal mass due to the self
  interactions~\eqref{eq:S^4}, $\alpha^2 = \dfrac{\lambda_{S}}{24}$
  (for $\lambda_{S}<1$).}
\begin{equation}
\mST^{2} \approx \mSV^{2}+ \alpha^{2} \; T^{2}\, .
\label{eq:mS}
\end{equation}
An important consequence of \eqs{eq:mS} is that the decay $S
\rightarrow \bar{\chi} \chi$ can become kinematically allowed at large
temperatures even if $\mSV<2\mc$. This feature will determine the
forbidden freeze-in regime.

Before discussing this regime it is worthwhile to note that in some models dark matter production at early times is dominated not by decays but by the $2 \leftrightarrow 2$ processes.\footnote{This is the case, \eg, for gravitino or axino dark matter at high $T_R$; for a review see, \eg~\cite{Baer:2014eja} and references therein.} In such cases the thermal effects typically introduce only a correction, the significance of which is very model-dependent.\footnote{Indeed, the dominant $2 \leftrightarrow 2$ process should involve the coupling between the mediator and the thermal bath, since it is assumed that it is in equilibrium contrary to the dark matter. A proper estimation of this effect can thus be done only on a model-dependent basis, as studied later in Sec.~\ref{sec:Higgsportal} for the Higgs portal case.} We will explicitly address the role of $2 \leftrightarrow 2$ production processes for the Higgs portal model in Sec.~\ref{sec:Higgsportal}, while for the following general discussion we restrict ourselves to the cases when they are subdominant.

As an example, let us consider the case $\mSV=0$, \ie, when $\mST = \alpha
T$. Assuming that the temperature is large enough so that at some
early time $\mST>2\mc$ is satisfied, our aim is to solve in this case the Boltzmann
equation~\eqref{eq:dYdx_std}. Defining $z$ as
\begin{align}
z \equiv \dfrac{2\mc}{\alpha  T} \, ,
\end{align}
we obtain
\begin{equation}
\dfrac{d\Yc}{dz}=\lrb{ \dfrac{\Gamma_\chi (m_S,m_\chi) }{ 5.93 \times 10^{-19} \, \GeV} }  \lrb{ \dfrac{1 \ \GeV}{2 \mc} }^2 \dfrac{\alpha^4 \; K_{1}(\alpha)}{ \sqrt{g}\; h } \delta_{h} \; z \, .
\label{eq:dYdx_pure_thermal}
\end{equation}

We observe two very different types of behavior depending on the
dimension $d$ of the operator that mediates the decay of $S$. In the
case when $d>4$, the right-hand side of \eqs{eq:dYdx_pure_thermal}
increases with temperature and is therefore dominant at high
temperatures close to the reheating temperature $T_{\rm R}$.  Thermal
effects in this case only provide a modification to the standard
freeze-in through higher-dimensional operators, as is the case for gravitino or 
axino DM produced in scatterings of particles in the thermal 
plasma~\cite{Roszkowski:2017nbc}. On the other hand, when $d \leq 4$  
most of the production takes place at temperatures around the dark matter mass. 
Indeed, DM production in this case increases at low temperature but stops when 
the decay becomes kinematically forbidden at $\alpha T = 2 \mc$. The production is 
thus dominated by temperatures close to $\mc$ (or higher for small $\alpha$).

In the following we will study in more details both cases to obtain a closed approximate form for the final dark matter aboundance when possible.

\subsubsection*{Higher-dimensional case}

 Let us first assume that $d>4$ in which case at high temperatures the thermal mass
of $S$ dominates and we can write its decay rate in the form
\begin{align}
\Gamma_\chi \sim \dfrac{\gamma_{S \chi}}{16 \pi} \; m_S \; \lrb{\frac{m_S}{\Lambda}}^{2n} = \dfrac{\gamma_{S \chi}}{16 \pi} \;\alpha^{2n+1}  \;
 \lrb{\frac{T}{\Lambda}}^{2n} T  \, ,
\label{eq:highTapprox}
\end{align}
where again $n=d-4$, and $\gamma_{S \chi}$ a dimensionless factor that depends 
on the nature of  this operator.  In the high-temperature regime where the 
approximation~\eqref{eq:highTapprox} is justified, the abundance equation becomes
\begin{equation}
\dfrac{d\Yc}{dz}= \lrb{ \dfrac{\gamma_{S \chi}  }{ 2.96 \times 10^{-17}} }  \lrb{ \dfrac{2 \mc}{\Lambda} }^{2n}  \lrb{ \dfrac{1 \ \GeV}{2 \mc} } \;  
\dfrac{\alpha^4 \; K_{1}(\alpha)}{ \sqrt{g}\; h } \delta_{h} \; z^{-2n} \, .
\label{eq:dYdxUV}
\end{equation}
Since the production is dominated by the high temperature contribution, it is
straightforward to integrate this equation, between $z=1$ (the decays are  
kinematically not allowed for $z \geq 1$) and $z = \zR \equiv \dfrac{2\mc}{\alpha \TR}$ 
to obtain
\begin{equation}
	\Ycr= 
	\frac{\zR^{1-2n} -1 }{2n-1}   \lrb{ \dfrac{\alpha^4 \; K_{1}(\alpha) \; \gamma_{S \chi}  }{ 2.96 \times 10^{-17} } }  \lrb{ \dfrac{2 \mc}{\Lambda} }^{2n}  
	\lrb{\dfrac{1 \ \GeV}{2 \mc} } \lrb{ \dfrac{1}{ \sqrt{g} \; h }  } \! \Big{|}_{z \sim \zR }\, .
	\label{eq:Y_frb_UV}
\end{equation}
It is clear that, for $d>4$ the dominant contribution comes from the
regime of high temperatures ($\zR \to 0$). An important consequence 
of the thermal effects included here is the fact that two-body decays 
can significantly alter the predictions of the scenario mentioned before 
(which was akin to the so-called ultraviolet freeze-in scenario advocated, 
\eg, in~\cite{Elahi:2014fsa}).
That is, even if the decays $S \to \bar{\chi} \chi$ are allowed in the vacuum, 
the appearance of the thermal mass of $S$ still plays a dominant role at high 
enough reheating temperature since in this case DM production is most efficient 
at high temperatures. Furthermore, comparing \eqs{eq:Y_std_UV} with \eqs{eq:Y_frb_UV}, 
we can see that   since $\alpha < 1$, the later tends to be generally less efficient.\footnote{ Usually  $2 \to 2$ processes involve higher powers of couplings, and  they are 
often subdominant to decays. For higher dimensional operators, however, both $1 \to 
2$ and $2 \to 2$ processes can involve similar powers of the couplings. This is the 
case where this argument is applicable.} Therefore, we conclude that the DM 
production via the forbidden freeze-in, in general, requires larger couplings in 
order to reproduce the observed relic abundance.

\subsubsection*{Four- or three-dimensional case} 

In the four (or three) dimensional case, most of the production is expected to  take 
place at low temperatures, as can be seen from \eqs{eq:Y_frb_UV} where the 
contribution from $z=z_{\rm R}$ drops out (unless $\alpha$ is so small  that the 
production happens close to the reheating temperature). 
More precisely, it takes place at around
the time when the decay $S\to \bar{\chi} \chi$ stops. Thus, we expect the production to be dominated at time scale corresponding 
to the temperature at which $\mST \sim 2 \mc$. This actually implies that up to an order  one function, the decay rate satisfies $\Gamma_\chi \propto m_\chi$.
While it is therefore not possible to fully simplify the decay rate without 
specifying the details of the interaction, we can straightforwardly observe from 
\eqs{eq:dYdxUV} that the abundance will be  proportional to $1/m_\chi$, thus 
implying that, up to order one corrections, the final relic density will be 
independent of the dark matter mass, as mentioned before.

As an example  and in order to obtain a closed form for the final relic density, 
let us assume that: $S$ is a scalar field, the dark matter
candidate $\chi$ is a Dirac fermion and the Lagrangian contains an
Yukawa interaction between $S$ and $\chi$,
\begin{equation}
\mathcal{L}_{\rm int}=- \, \ysx    \, \bar{\chi}\chi \, S \, .
\label{eq:yukawa}
\end{equation}
The bath particle  $S$  decay width to dark matter is then given by
\begin{equation}
\Gamma_{S \to \bar{\chi} \chi}  = \frac{\ysx^{2}}{8 \pi} \dfrac{\lrb{  m_{S}^{2}- 4m_{\chi}^{2} }^{3/2} }{m_{S}^{2}} \, .
\label{eq:GammaS}
\end{equation}
The evolution of the yield is then given by
\begin{equation}
\dfrac{d\Yc}{dz} = \lrb{\dfrac{\alpha^2 \; \ysx }{3.86 \times 10^{-9}}}^{2} \lrb{ \dfrac{1~\GeV}{2\mc} } \; K_{1}(\alpha) \; \dfrac{ \lrb{ 1-z^{2} }^{3/2} }{ \sqrt{g}\; h } \delta_{h} \, .
\label{eq:dYdx_Sxx}
\end{equation}
In Fig.~\ref{fig:Y_ev_frb} we show the evolution $\Yc/ \Ycr$
of the number of DM particles as a function of $z$ for the thermal
mass case. It is similar to the standard case, apart from the point
when the production stops, \ie, at $\mST = 2\mc$.

Assuming that the relativistic degrees of freedom do not vary rapidly
during the production of the $\chi$s, 
we can integrate \eqs{eq:dYdx_Sxx} to obtain
\begin{equation}
\Ycr= \lrb{\dfrac{\alpha^2 \; \ysx }{ 5 \times 10^{-9}}}^{2} \lrb{ \dfrac{1 ~\GeV}{2\mc} } \; K_{1}(\alpha) \;  \lrb{ \dfrac{  1 }{ \sqrt{g}\; h } }_{z=\vev{z}}  \, ,
\label{eq:Y0_Sxx}
\end{equation}
where again $g$ and $h$ are evaluated at $\vev{z}$.\footnote{In this case, $\vev{z}$ is defined as
\begin{equation*}
\vev{z}\equiv \dfrac{\int_{0}^{1} dz \; \lrb{ 1-z^{2} }^{3/2} \times z }{\int_{0}^{1} dz \,\lrb{ 1-z^{2} }^{3/2} }  \approx 0.34 \, .
\end{equation*}}
As we pointed out earlier, the relic abundance of $\chi$ becomes (mostly) 
independent of its mass, with any $\mc$ dependence coming from $\vev{z}$.
This, and the suppression due to the $\alpha^{4}$, will result in relaxed 
constraints for the Yukawa coupling, with respect to the standard freeze-in, 
where $\omh$ scales predominantly linearly with the DM mass. 
Notice furthermore that in the case where the temperature correction never 
dominates (\ie \, $\alpha \, T<\mSV$),  the relic abundance is given 
by \eqs{eq:relic_def} with the  decay width~\eqref{eq:GammaS}, which is the 
standard freeze-in case, as expected. 

Finally, let us conclude this section by presenting some numerical
results in the case where both $\mSV$ and $\mST$ play an important role
as the temperature varies.  In this case one has to calculate $\Ycr$ by
including both mass terms. That is, the evolution of $\Yc$ as in
\eqs{eq:dYdT_std} needs to be solved, with $\mST$ given
by~\eqs{eq:mS}, numerically.
%%%%%%%%%%%%%%%%%%%%%%%%%%%%%%%%%%%%%%%%%%%%%%%%%%%%%%%
\begin{figure}[t]
	\centering
	\begin{subfigure}[b]{0.49\textwidth}
		\centering
		\includegraphics[width=1\textwidth]{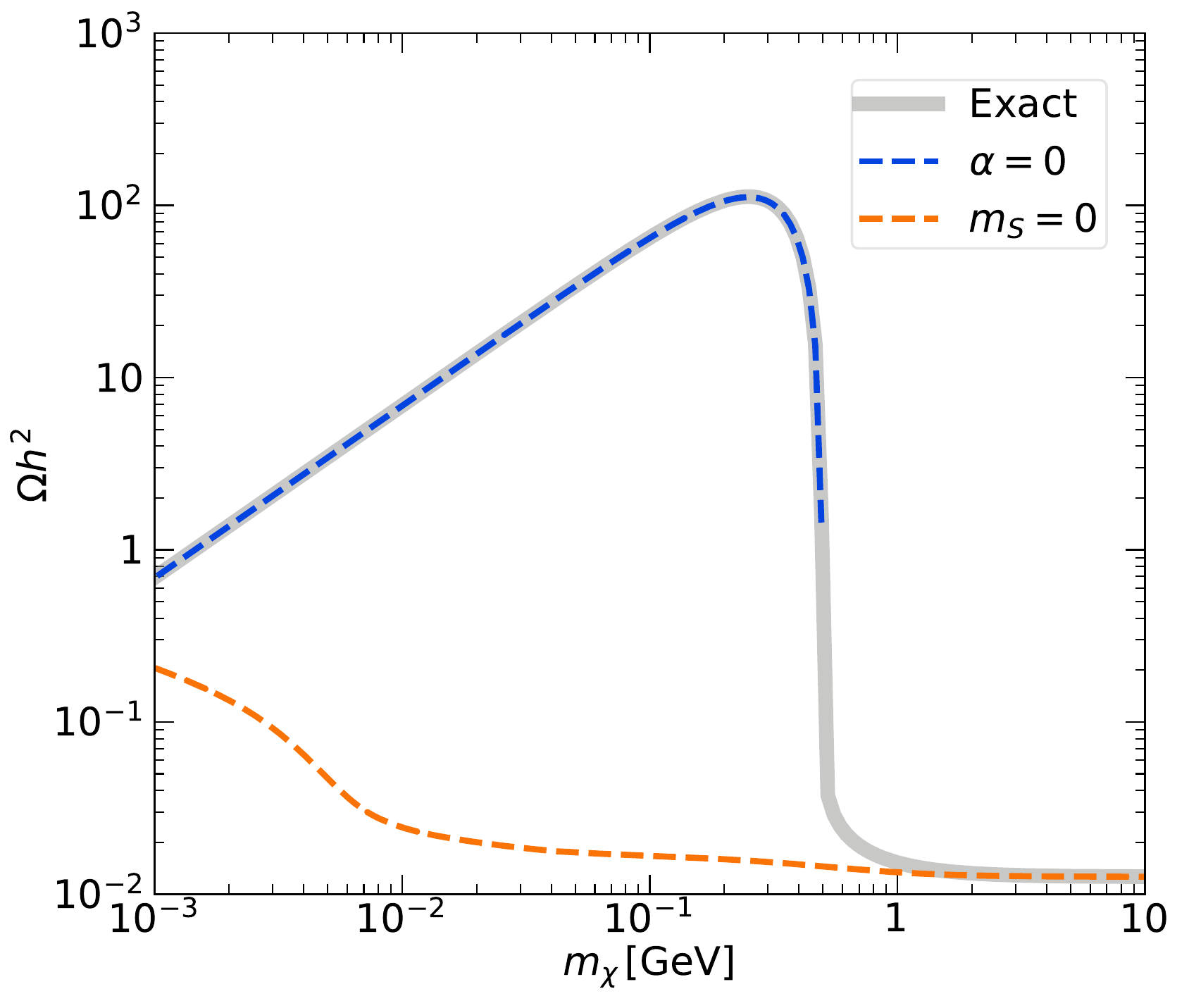}
		\caption{}
		\label{fig:limits}
	\end{subfigure}
%	\hfill
	\begin{subfigure}[b]{0.49\textwidth}
		\centering
		\includegraphics[width=1\textwidth]{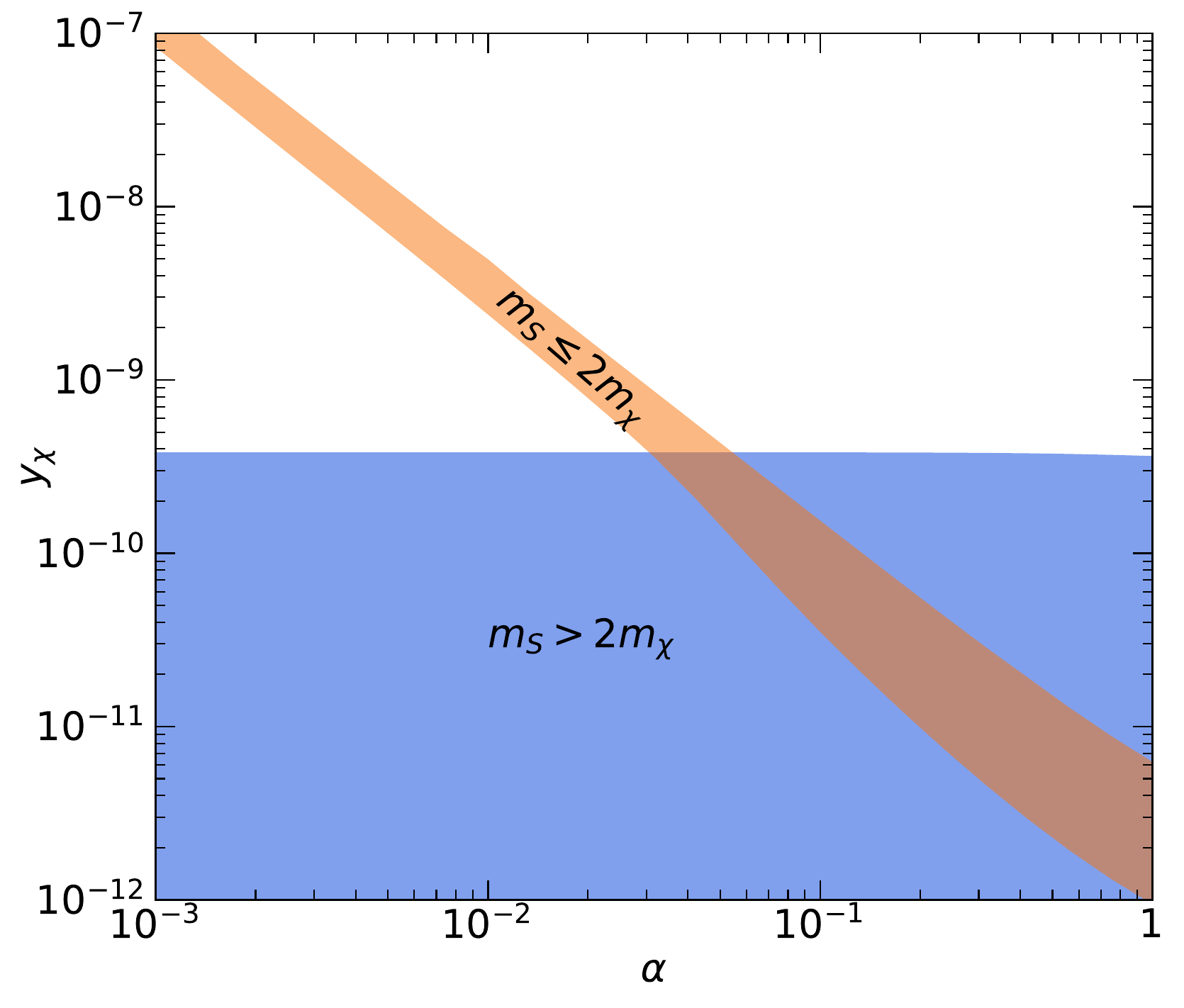}
		\caption{}
		\label{fig:scan}
	\end{subfigure}
	\caption{{\bf (a)} The relic abundance for $\mSV=1\; \GeV$,
		$\alpha=0.1$, and $\ysx =5 \times 10^{-11}$.  The exact
		result is shown in gray, while the other lines correspond
		to the limits of dominant (blue) and vanishing (orange)
		$\mSV$ .  {\bf (b)} The area in the plane $\alpha-\ysx $,
		where the observed relic can be obtained for $ 10 \; \MeV
		\leq \mSV,m_{\chi} \leq 1\; \TeV $. The two shaded regions
		correspond to the forbidden freeze-in region $\mSV < 2m_{\chi}$ (orange) and
		the standard one $\mSV>2m_{\chi} $ (blue).}
%	\label{fig:limits}
\end{figure}

%%%%%%%%%%%%%%%%%%%%%%%%%%%%%%%%%%%%%%%%%%%%%%%%%%%%%%%%%%%
An example of typical dependence of $\omh$ on $\mc$ for the production of 
DM due to the decay of $S$, is shown in Fig.~\ref{fig:limits}. The two 
extreme cases of $\alpha = 0$ (standard freeze-in) and $\mSV =0$ 
(dominance of the thermal corrections to the mass) are shown by dashed 
blue and orange lines, respectively, while the exact numerical result is 
shown in solid grey. Notice that the transition between the two limits happens
suddenly at $\mc \approx \mSV / 2$ which is where the blue line
terminates since $S \to \bar{\chi} \chi$ becomes forbidden in the vacuum.

%%%%%%%%%%%%%%%%%%%%%%%%%%%%%%%%%%%%%%%%%%%%%%%%%%%%%%%%%%%

In Fig.~\ref{fig:scan} we present the Yukawa coupling $\ysx$ as a
function of $\alpha$ that give the observed $\omh$ for the scanned
range of masses $ 10 \; \MeV \leq \mSV,\mc \leq 1\; \TeV $, hence
overlapping regions between the two regimes may correspond to
completely different values of the masses. We observe two distinct
regimes: the region of standard freeze-in where $\mSV > 2m_{\chi}$ is
marked in blue, while the forbidden freeze-in region of $\mSV <
2m_{\chi}$ is marked in orange. The shape of the forbidden freeze-in
band in Fig.~\ref{fig:scan} is a simple consequence of the
$\alpha^2\ysx$ dependence of $\Ycr$ in \eqs{eq:Y0_Sxx}. As already
noted in the $d>4$ case, in the forbidden freeze-in regime one
requires either larger self-interaction $\alpha$ of the mediator to
generate a larger thermal mass, or a stronger interaction coupling
between DM and the mediator, since the DM production is not as
efficient as the standard case (as also shown in
Fig.~\ref{fig:limits}). An important comment is that the transition
between the two regimes, which happens for $\mSV \sim 2m_{\chi}$
occurs typically in a mass range of order $(2m_{\chi} - \mSV) \sim
\alpha \mSV$, which become very narrow for small
$\alpha$.\footnote{This corresponds to the case where the mass
  difference preventing the decay of $S$ into two DM particles
  is of the same order as the thermal contribution to $\mSV$ at the
  typical scale $T \sim \mSV$. In particular, in Figure~\ref{fig:scan} the forbidden region shown in orange do not probe this tuned transition regime in details for small $\alpha$.}

\section{Forbidden freeze-in and the Higgs portal}\label{sec:DarkHiggs}
\label{sec:Higgsportal}
\setcounter{equation}{0}
In this section we explore an explicit realisation of the general mechanism 
described above. We focus on a Higgs portal model, which is an archetype for a wide 
class of DM models where the dark sector is connected to the visible sector by a 
scalar mediator mixing with the SM Higgs boson. 

\subsection{The model}\label{sec:Model}

We introduce a real scalar ``dark Higgs'' boson field $S$, which is
not protected by a $\mathcal{Z}_{2}$ symmetry and hence can decay into
Standard Model fields through its mixing with the SM Higgs boson.  A
dark matter candidate is taken to be a Dirac fermion that couples to
the dark Higgs boson through a small Yukawa coupling $\ysx$. The
corresponding part of the Lagrangian thus reads
\begin{align}
\mathcal{L}^{\rm DM} & =\bar{\chi}\lrb{ i \gamma_\mu D^\mu-\mu_{\chi}  } \chi +  \frac{1}{2} (D^\mu S ) (D_\mu S)  
 -\ysx    S \bar{\chi}  \chi  -  V_{HS}       \, , 
\end{align}
with the dark Higgs boson potential term defined as~\footnote{Notice
  that several other operators can be written within our symmetries,
  including a trilinear coupling $S^3$ and Yukawa couplings to left
  and right components of the dark matter fermion. We will neglect the
  trilinear in the following and enforce an exact $\chi$-number global
  symmetry to fix the latter to zero.}
\begin{equation}
	V_{HS}= \dfrac{\mu_{S}^{2}}{2} \, S^2 +  \dfrac{\lambda_S}{4!} \, S^4 + 
	A \, S \, H^{\dagger} H + \lambda_{HS} \, S^2 \, H^{\dagger} H \, ,
	\label{eq:VHS}
\end{equation}
where $H$ denotes the Standard Model Higgs boson doublet. The total scalar potential 
is $V=V_{HS}-\mu  \, H^{\dagger} H + \dfrac{\lambda_{H}}{2} \, \lrb{H^{\dagger} 
H}^{2} $.

At low temperatures ($T \lesssim 160 \; \GeV$), both the Higgs and
dark Higgs fields develop a non-zero vacuum expectation value (VEV),
so that $H=\dfrac{1}{\sqrt{2}}\lrb{ \begin{matrix} 0  \\
    h+v \end{matrix} }$ and $S \to v_S + S$.~\footnote{Since $S$ plays
  a crucial role in the production of DM before and after EW phase
  transition, we just denote the VEV-shifted dark Higgs boson as $S$
  in order to avoid changing the notation when dealing with different
  temperature regimes.}
In the limit where $ A\ll v$ the calculation simplifies significantly 
and the minimization conditions for the scalar potential in term of $\lambda_{H}$ 
and $v_S$ can be easily obtained as
\begin{align}
\lambda_{H} \approx &  \lrb{ \dfrac{2 \mu_{H}}{v} }^{2}  +   \lrb{ \dfrac{A}{\mSV} }^{2}  \nonumber \\
v_S \approx & - \frac{A v^2}{2 \mSV^2 } \, .
\label{eq:minimization_approx}
\end{align}
Furthermore, we can rotate the scalars to their eigenvalue basis, \ie $\lrb{ \begin{matrix} h , S \end{matrix} } \to R \lrb{ \begin{matrix} h , S \end{matrix} }$,
where $R$ is a rotation matrix parametrised by the small angle   $\theta$ given by
\begin{equation}
\theta =  \frac{A \, v }{m_{h}^{2} - \mSV^{2}} \lrsb{ 1 -   \frac{\lambda_{HS} \; v^{2}}{\mSV^{2}}  } \, .
\label{eq:theta}
\end{equation}
where we have used the masses of $h$ and $S$ (at $T=0$) defined by
\begin{align}
m_{h}^{2} =  \lambda_{H} v^{2} \ \text{and} \  
\mSV^{2} =  \mu_{S}^{2} + \lambda_{HS} v^{2}  \, .
\label{eq:masses}
\end{align}
The branching ratio of the Higgs decay to invisible particles is
constrained to be~\cite{Belanger:2013xza} smaller than $0.19$, which
translates to $\lambda_{HS}\lesssim 10^{-2}$. Furthermore, note that
while we have supposed that the trilinear term $\lambda_{3} S^{3}$ was
negligible in our original Lagrangian,\footnote{For example, this term
  can shift the thermal mass of $S$ by a factor of $\mathcal{O} \lrsb{
    \lambda_S \lrb{\dfrac{v_{S}}{\mu_{S}}}^{2}}$.} the shift by
$v_{S}$ re-introduces such a term as $ \dfrac{\lambda_S}{3!} v_{S}
S^{3}$. For consistency, we will therefore further require that this
contribution is negligible with respect to $\mu_S$, leading to the
condition
\begin{align}
\frac{A}{\mu_S} \ll \frac{ 12 \mSV^2}{\lambda_S v^2} \, .
\label{eq:smalltri}
\end{align}
Notice that this also automatically ensures that the shift in the SM Higgs boson 
quartic coupling $\lambda_H$ is negligible in \eqs{eq:minimization_approx}.  An 
interesting feature is that the dark Higgs boson is extremely long-lived at low 
mass. When only its decays into a lepton $\ell$ pair are kinematically allowed, and
assuming $\mu_S \sim \mSV$, we obtain
\begin{align}
	\tau_S = \frac{8 \pi \hbar}{\mSV y_\ell^2 \theta^2} ~\gg~ \begin{cases}
	4 \cdot 10^6 \textrm{ s } \times \lambda_S^2  \displaystyle \lrb{\frac{100 \textrm{ MeV}}{\mSV}}^7 \quad \textrm{ for $S \rightarrow e^+e^-$} \, ,\\[1em]
	0.15  \textrm{ s } \times \displaystyle \lambda_S^2
        \lrb{\frac{250 \textrm{ MeV}}{\mSV}}^7 \quad \textrm{ for $S
          \rightarrow \mu^+\mu^-$} \, .	\end{cases}
	\label{eq:lifetime}
\end{align}
As we will see in the next section, such long lifetime are severely
constrained by astrophysical limits and beam dump limits. For
simplicity, we will therefore typically restrict ourselves to $\mSV >
100$ MeV in the following.\footnote{Note, though, that strictly
  speaking one could still satisfy the above bounds while keeping the
  $\tau_S \sim 0.1$~s, for very low values of $\lambda_S$. The
  parameter space is however extremely restricted experimentally, as
  we will see in Sec.~\ref{sec:explimits}.}

\begin{figure}[t]
	\centering
	\begin{subfigure}[b]{0.29\textwidth}
		\centering
		\includegraphics[width=1\textwidth]{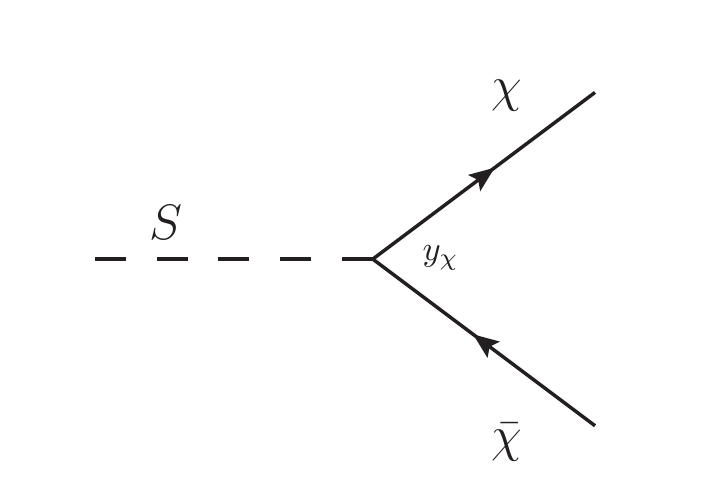}
		\caption{}
		\label{fig:diag1}
	\end{subfigure}
	\hspace{-1.cm}\begin{subfigure}[b]{0.29\textwidth}
		\centering
		\includegraphics[width=1\textwidth]{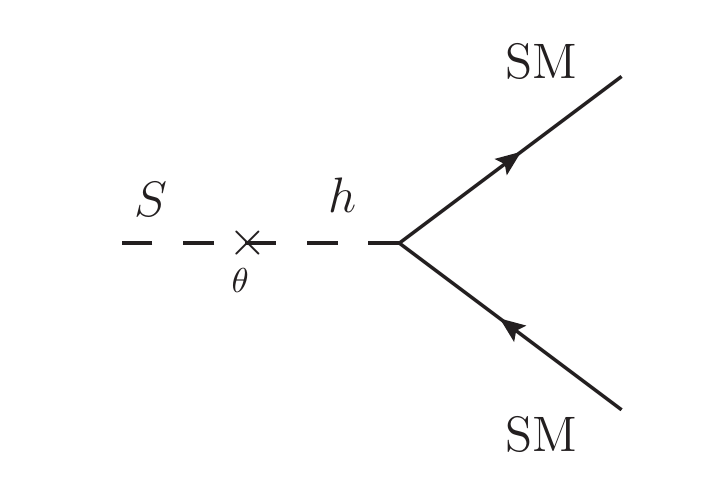}
		\caption{}
		\label{fig:diag2}
	\end{subfigure}
	\hspace{0.5cm}\begin{subfigure}[b]{0.29\textwidth}
		\centering
		\includegraphics[width=1\textwidth]{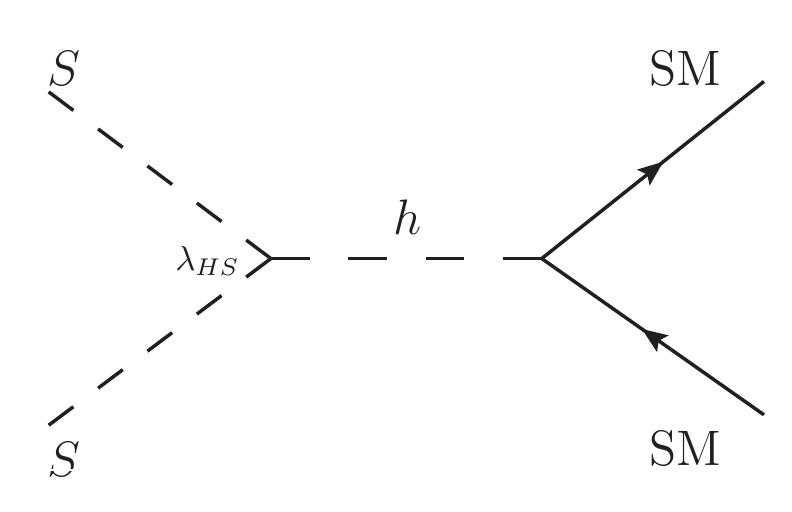}
		\caption{}
		\label{fig:diag3}
	\end{subfigure}
	\caption{Feynman diagrams for the dominant processes governing
		the freeze-in of $\chi$. {\bf (a)} The main $\chi$
		production mode through mediator decay. {\bf (b)} The
		mediator decay to SM particles through mixing with the Higgs
		affecting both the freeze-out of $S$ and the branching
		fraction of late time decays. {\bf (c)} Pair annihilation of
		$S$ contributing to the freeze-out of $S$. The relative
		importance of these processes is to large extent determined
		by the hierarchy of the highlighted couplings $y$ and
		$\lambda_{HS}$ as well as by the mixing angle $\theta$.}
	\label{fig:diagrams}
\end{figure}

The relevant processes determining the evolution of number densities
of $S$ and $\chi$ in this model are: \textit{i)} the direct mediator
decay $S\rightarrow \bar\chi\chi$, \textit{ii)} the mediator decay to
SM particles due to its mixing with the SM Higgs boson, and
\textit{iii)} the annihilation of $S$ to SM particles, as well as all
the inverse reactions. The Feynman diagrams for these processes are
given in Fig.~\ref{fig:diagrams}.\footnote{Note that for heavy
  mediators a decay/annihilation channels to $h$ could be open
  resulting in additional two processes $S\rightarrow hh$ and
  $SS\rightarrow hh$, governed by $A$ and $\lambda_{SH}$,
  respectively. These are not relevant in our analysis, which is
  focused on the regime $\mSV<m_h$ for the temperatures around the
  freeze-out temperature of $S$, and therefore are not included.} The
direct $S \to \bar{\chi} \chi$ decay width is given by \eqs{eq:GammaS}
and is suppressed by the very small Yukawa coupling $\ysx$. The decay
of $S$ to SM particles is given by $\Gamma(S\rightarrow {\rm SM}) =
\theta^2\; \Gamma_{h\to {\rm{SM}}}(\mST)$, where the $\Gamma_{h\to
  \rm{SM}}(\mST)$ is the total width of the SM-like Higgs boson with
mass $\mST$. We implement using the results taken from
\cite{Dittmaier:2011ti,Fradette:2017sdd,Winkler:2018qyg} and a direct
evaluation for leptonic decay at low masses.

The $S$ annihilation cross section as a function of the Mandelstam variable $s$ reads
\begin{equation}
\sigma v (SS\to  {\rm{SM}} )= \frac{\Gamma_{h\to \rm{SM}}(\sqrt{s})
}{\sqrt{s}} \,\frac{8\lambda_{HS}^2 v^2}{(s-m_h^2)^2+m_h^2 \Gamma_h^2
} \, .
\label{eq:sigmav}
\end{equation}
It can be of the order of the standard WIMP annihilation
cross-section, or smaller. This is because $S$ is unstable and therefore its
number density right after freeze-out can be much larger than for
standard WIMP. We will assume that either $\lambda_{HS}$ or the mixing
angle $\theta$ are large enough to ensure that $S$ was in equilibrium
at very early times (see discussion in the next section).

Apart from the processes shown in Fig.~\ref{fig:diagrams}, additional
$2\leftrightarrow 2$ processes can in principle play a role in the
production of $\chi$s and/or their early-time thermalization with the
SM plasma. These are: $SS\leftrightarrow \bar\chi\chi$,
$hh\leftrightarrow \bar\chi\chi$ and the co-annihilation process
$Sh\leftrightarrow \bar\chi\chi$. The first one has $s-$, $t-$ and
$u-$channel contributions which are proportional to
$\theta^2\lambda_{HS}^2 \ysx^2$ and $\ysx^4$, respectively. The second
and third have only $s-$channel diagrams proportional to $A^2\ysx^2$
and $\lambda_{HS}^2 \ysx^2$, respectively. It is clear that all of
theses $2\leftrightarrow 2$ processes are strongly suppressed with
respect to direct $S$ decays due to phase space suppression exhibited
by the $2-$body phase space of the former channels.
However, in the deeply forbidden regime (\ie, for very small
$\lambda_{S}$), when the decay is kinematically allowed only at very
high temperatures, all the aforementioned channels could in principle
play some role in the evolution of $\chi$.  In light of this, we have
implemented all of the above processes in the numerical approach
presented in the next section and checked explicitly that for the
parameter ranges covered by our scan these processes indeed can be
safely neglected in solving the evolution equations of $S$ and $\chi$
number densities.

\subsection{Relic density and numerical study}

In light of the above discussion the coupled computation of the
freeze-out of $S$ and the freeze-in of $\chi$ is performed under the
assumptions that: \textit{i)} $\chi$ had negligible abundance after
reheating and had not reached chemical equilibrium, \textit{ii)} $S$
was in chemical equilibrium at early times and remained in kinetic
equilibrium for all the temperatures relevant for the production of
$\chi$s.\footnote{Kinetic equilibrium is an extremely good assumption
	in the parameter space studied in this work since away from the
	Higgs boson resonance elastic scatterings of $S$ off particles of
	the SM plasma are much more frequent than annihilations of $S$. In a
	different model where this assumption would be violated one would be
	required to solve also for the temperature of $S$ or even its full
	phase space density, see~\cite{Binder:2017rgn}. This would also
	bring additional complication to the forbidden freeze-in case as the
	thermal mass of $S$ would need to be computed out of equilibrium. In
	fact, even if $S$ is still in kinetic equilibrium (with the SM
	plasma or with itself), but already chemically frozen-out, the
	thermal mass would not be given by \eqs{eq:mST}. However, this
	caveat has no implications for our results since in the studied
	model the forbidden freeze-in happens at large enough temperatures
	where $S$ is still in equilibrium. }  In practice, the
assumption made in the numerical code is that  the above conditions
are satisfied up to $x=0.1$, where we define $x\equiv
\mSV/T$. For $x<0.1$ it is assumed that $S$ traces its equilibrium
value while the evolution of $\chi$ is given by \eqs{eq:dYdx_std},
starting from the reheating temperature $\TR$ assumed to be given by
$\xR=10^{-9}$. We checked explicitly that assuming different
$\TR$ does not change the result. For $x>0.1$ the coupled system of
the Boltzmann equations for the number densities of $S$ and $\chi$ is
numerically solved, including all the relevant processes discussed
above.\footnote{This is done to ensure that the $\chi$ production from
	$S$ decay takes into account possible deviations from chemical
	equilibrium of $S$. As stated before, this does not affect the
	forbidden freeze-in regime in our model, but it does some part of
	the parameter space of the standard freeze-in. For discussion and
	explicit forms of suitable Boltzmann equations see
	\eg~\cite{Belanger:2018mqt}.}

Within this setup there are several possible regimes leading to the
correct DM abundance. In the following we first show some representative
examples of the evolution of the yields of $S$ and $\chi$ for different
regimes and then present and discuss the results of our scan of the
parameter space of the model.

\subsubsection{Evolution of number densities}
In Figures~\ref{fig:Y1}--\ref{fig:Y3} we present the yields of $S$ and $\chi$ for some
characteristic cases. In all following figures the green dashed lines
correspond to $Y_S$ while the solid lines to $Y_{\rm DM}$ with the
blue color indicating standard (non-forbidden) regimes and the beige
one forbidden regimes. For completeness, the light gray area
highlights the evolution of the yields during the time before the
electroweak phase transition (EWPT). In all the plots the different
shadings of the lines correspond to the variation of the most relevant
parameter for a given regime, as indicated in the figures.

The simplest case is the usual freeze-in, where $\mSV>2m_\chi$ and
$Y_{\rm DM}$ gradually grows, with most of the production happening
around $T\sim \mSV$. This is shown in Fig.~\ref{fig:Fig1a}. In this
case the final relic abundance of $\chi$ is insensitive to any
variations in the self-coupling $\lambda_S$ due to the fact that the
thermal effects are important only for $T\gg \mSV$, which is a very
short (in real time) period. Thus, the thermal mass of $S$ has a very
small impact on the result in the standard freeze-in
regime, as expected. Additionally, note that the equilibrium number density of $S$
is also affected only at early times due to thermal corrections, as they
shift the value of $\mST$.

In Fig.~\ref{fig:Fig1b} we show a typical case of forbidden freeze-in, where
an opposite behaviour can be seen. The production is active only at
small $x$ and is both stronger and terminates later for larger values
of $\lambda_S$. In this forbidden regime the final DM abundance is
therefore very sensitive not only to value of $\ysx$ but also the
self-coupling of the mediator. Another point worth stressing is that
one does not need large values of $\lambda_S$ to get a sizable effect,
so the opening of the forbidden decay due to thermal effects is in
fact a generic feature of the freeze-in mechanism.

\begin{figure}[t]
	\centering
	\begin{subfigure}[b]{0.49\textwidth}
		\centering
		\includegraphics[width=1\textwidth]{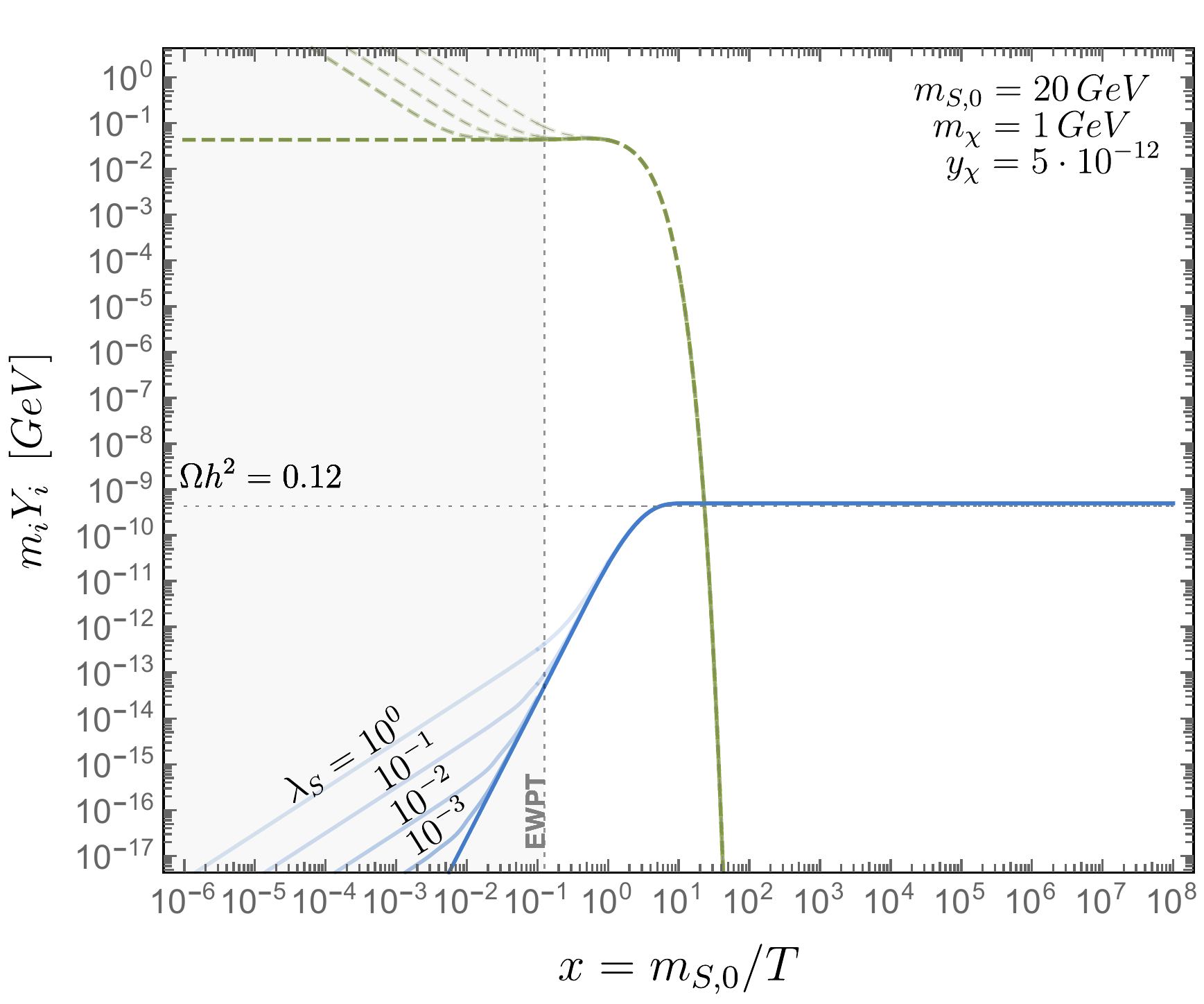}
		\caption{}
		\label{fig:Fig1a}
	\end{subfigure}
	%	\hfill
	\begin{subfigure}[b]{0.49\textwidth}
		\centering
		\includegraphics[width=1\textwidth]{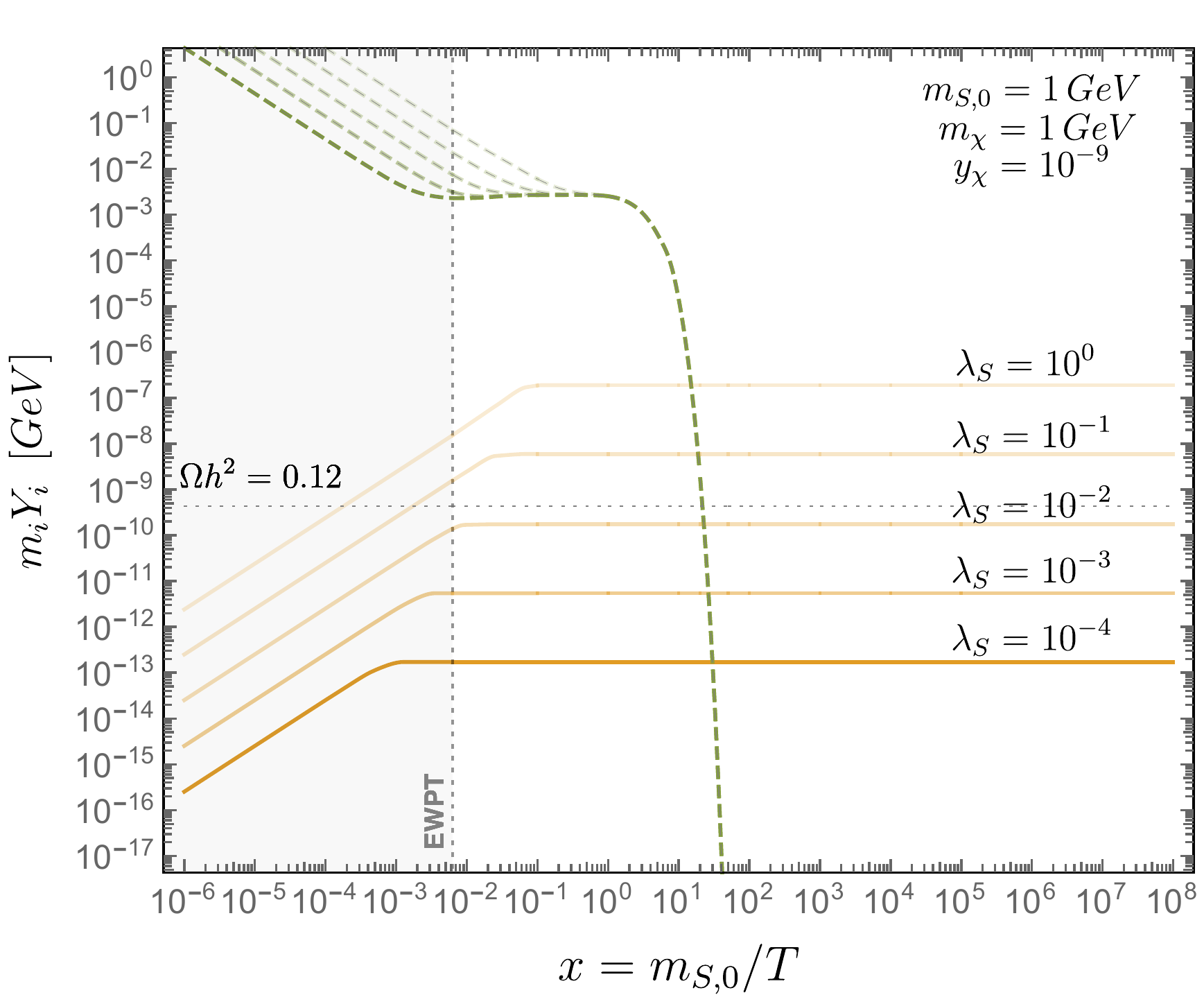}
		\caption{}
		\label{fig:Fig1b}
	\end{subfigure}
	\caption{Typical evolution of the yields of $S$ (dashed green)
		and $\chi$ (solid). The lower the line opacity the larger
		the self-coupling $\lambda_S$. {\bf (a)} A standard
		freeze-in case where the impact of $\lambda_S$ on the yields is only
		important at very high $T$ when there is not enough time to
		produce significant amounts of $\chi$ particles, leading to
		approximately the same value of their final relic abundance. {\bf (b)} A
		forbidden freeze-in case where the thermal mass of $S$ has the
		dominant effect that opens up $\chi$ production, hence one
		finds a  very strong dependence of $\Omega h^2$ on the self-coupling
		$\lambda_S$. }
	\label{fig:Y1}
\end{figure}

%
%\begin{figure}[t]
%	\centering
%	\includegraphics[width=0.49\textwidth]{figs/Fig1a.pdf}
%	\includegraphics[width=0.49\textwidth]{figs/Fig1b.pdf}
%	\caption{Typical evolution of yields for $S$ (dashed green) and $\chi$ (solid). The lower the line opacity the larger the self coupling $\lambda_S$. \textit{Left:} the standard freeze-in regime - the impact of $\lambda_S$ is only important at very high $T$, when there is not enough time to produce significant amounts of $\chi$ particles, leading to approximately the same final relic abundance. \textit{Right:} the forbidden freeze-in regime - the thermal mass of $S$ is the dominant effect that allows for $\chi$ production, therefore very strong dependence of $\Omega h^2$ on the self-coupling $\lambda_S$.} 
%	\label{fig:Y1}
%\end{figure}

Figure~\ref{fig:Fig2a} shows a case of a transition between the
standard and the forbidden regimes. For fixed $\mSV=100$~GeV we vary
$m_\chi$ and see that, as expected, around the transition the result
is very sensitive to precise value of the DM mass. In the forbidden
regime increasing $m_\chi$ further leads to only very mild change in
the relic abundance, \ie, the yield $Y_{\rm DM}$ is inversely
proportional to $m_\chi$, in agreement with \eqs{eq:Y0_Sxx}. This
approximate DM mass independence of the relic density is an distinct
feature of the forbidden freeze-in scenario.

In Fig.~\ref{fig:Fig2b} a slightly different mechanism is shown. It
occurs when nominally this would be a standard freeze-in case with
$\mSV>2m_\chi$ but, due to the EWPT and its effect on the mass of $S$
(which arises when the SM Higgs gets its VEV due to the presence of
the mixing quartic coupling $\lambda_{HS}$), there appears a temporary
regime where $S\to \bar\chi\chi$ is not allowed and the $\chi$
production is blocked for a while. However, if the self-coupling
$\lambda_S$ is large enough the thermal mass overcomes the suppression
due to the EWPT and re-opens the decay. This is an example of a
situation when the thermal mass has a large impact on the relic
abundance even in the standard freeze-in regime of $\mSV>2m_\chi$. A
scenario like this is close to what was studied, in a more general
context, in ref.~\cite{Baker:2017zwx}.

\begin{figure}[t]
	\centering
	\begin{subfigure}[b]{0.49\textwidth}
		\centering
		\includegraphics[width=1\textwidth]{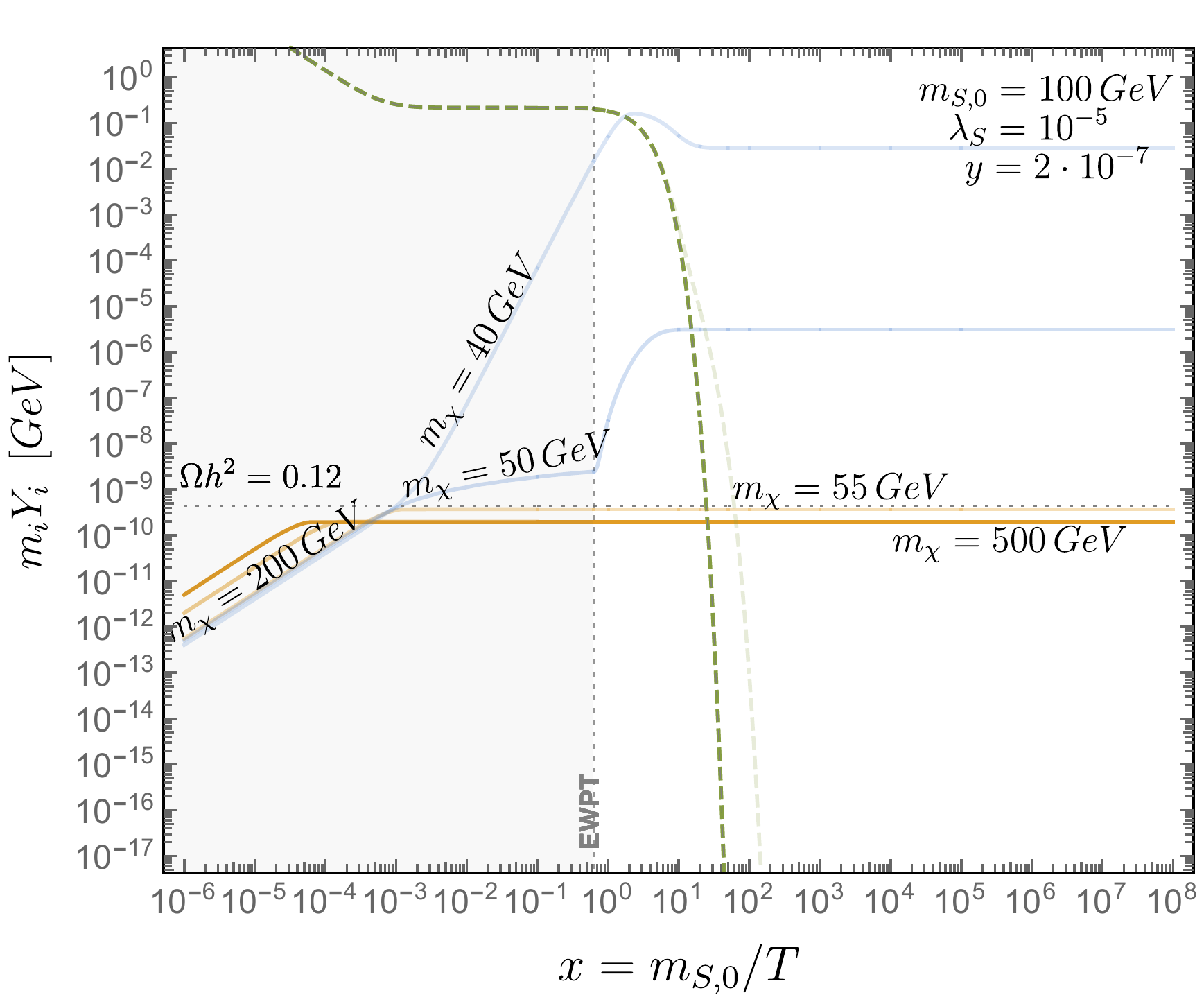}
		\caption{}
		\label{fig:Fig2a}
	\end{subfigure}
	%	\hfill
	\begin{subfigure}[b]{0.49\textwidth}
		\centering
		\includegraphics[width=1\textwidth]{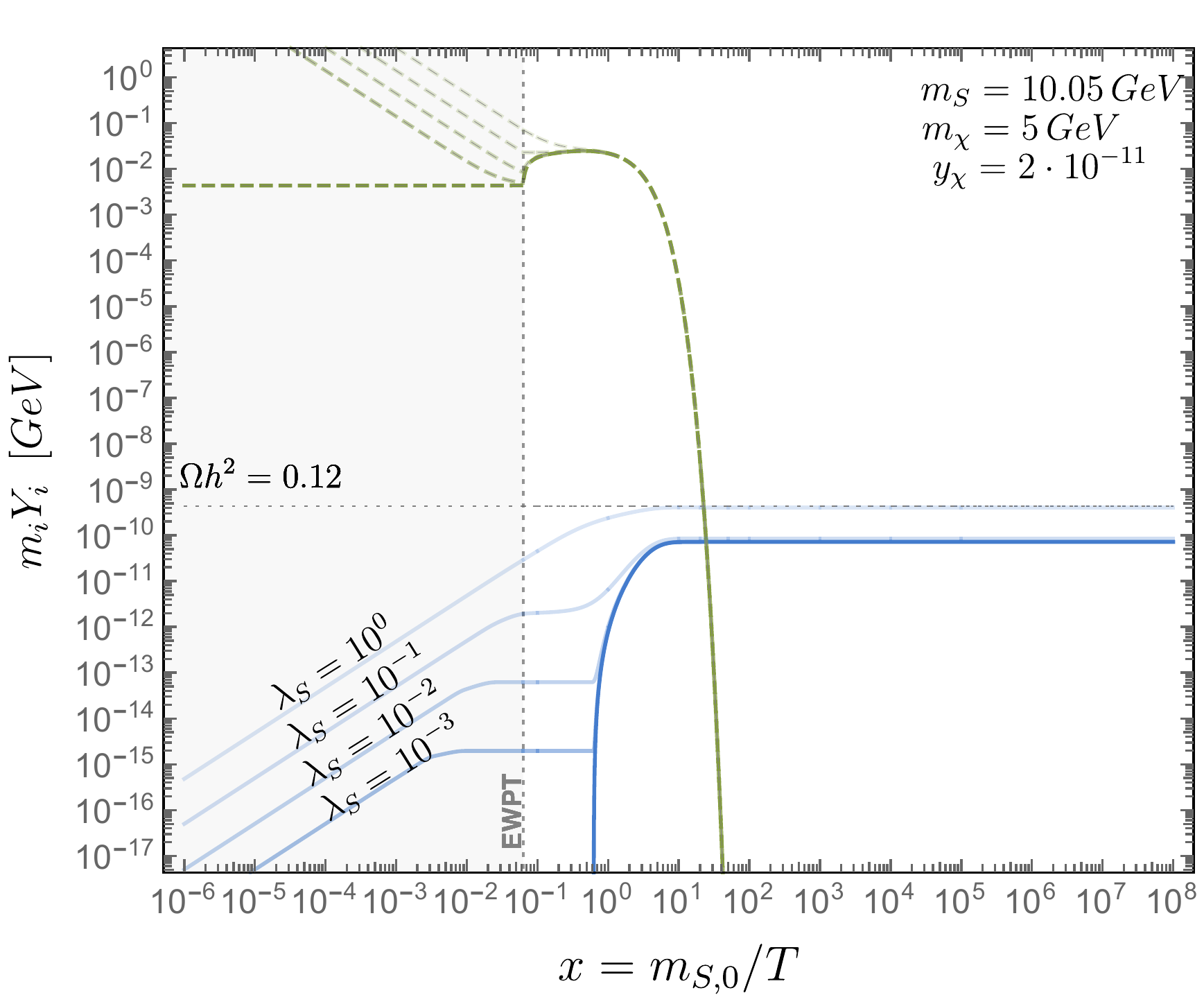}
		\caption{}
		\label{fig:Fig2b}
	\end{subfigure}
	\caption{ {\bf (a)} A transition between the standard and the
		forbidden freeze-in regimes. In the former (blue solid
		lines) the final abundance depends strongly on the $m_\chi$,
		while after a sharp transition to the forbidden regime
		$\Omega h^2$ is only very mildly dependent on the DM
		mass. {\bf (b)} Around the EWPT the $T$-dependence of the
		VEV causes a temporary regime where in the standard case the
		$S\rightarrow \bar\chi \chi$ is forbidden and $\chi$
		production is blocked. However, if the self-coupling
		$\lambda_S$ is large enough the thermal mass overcomes the
		suppression of $\mST$ due to the EWPT and re-opens the
		decay.}
	\label{fig:Y2}
\end{figure}

%\begin{figure}[t]
%	\centering
%	\includegraphics[width=0.49\textwidth]{figs/Fig2a.pdf}
%	\includegraphics[width=0.49\textwidth]{figs/Fig2b.pdf}
%	\caption{ \textit{Left:} the transition between standard and forbidden freeze-in regimes. In the former (blue solid lines) the final abundance depends strongly on the $m_\chi$, while after sharp transition to the forbidden regime the $\Omega h^2$ is only very mildly dependent on the DM mass. \textit{Right:} around the EWPT the $T$-dependence of the VEV causes a temporary regime where the $S\rightarrow \bar\chi \chi$ is forbidden and $\chi$ production is blocked. However, if the self-coupling $\lambda_S$ is large enough the thermal mass overcomes the suppression of $\mST$ due to the EWPT and re-opens the decay.}  
%	\label{fig:Y2}
%\end{figure}

Finally, in Fig.~\ref{fig:Y3} we show for completeness examples of
cases where the $\chi$ production is dominated by the late-time decay
of $S$. These cases are not directly related to the main focus of this
work but are present in some regions of the parameter when we scan of
the full model and therefore important in their own right. In these
cases the complete evolution of both $S$ and $\chi$ is crucial. In
Fig.~\ref{fig:Fig2a} the final DM abundance is determined by the
branching fraction of the $S$ decays to $\chi$ and to SM particles
which in the plot is parametrised by the value of the trilinear
coupling $A$. For smaller values (corresponding to a weaker mixing
with the SM Higgs boson), DM particles constitute a larger fraction of
$S$-decay products.

Figure~\ref{fig:Fig3b} shows a situation where the details of the
freeze-out of $S$ strongly affect its abundance that is then
transferred to the $\chi$s via (rare) decays. This also shows the
potential impact that the choice of $\lambda_{HS}$ can have on the
final relic abundance of DM. Note that in this plot different lines
correspond to different relation between $x$ and $T$ due to electroweak
symmetry breaking contribution to $m_S$ which depends on $\lambda_{HS}$.

\begin{figure}[t]
	\centering
	\begin{subfigure}[b]{0.49\textwidth}
		\centering
		\includegraphics[width=1\textwidth]{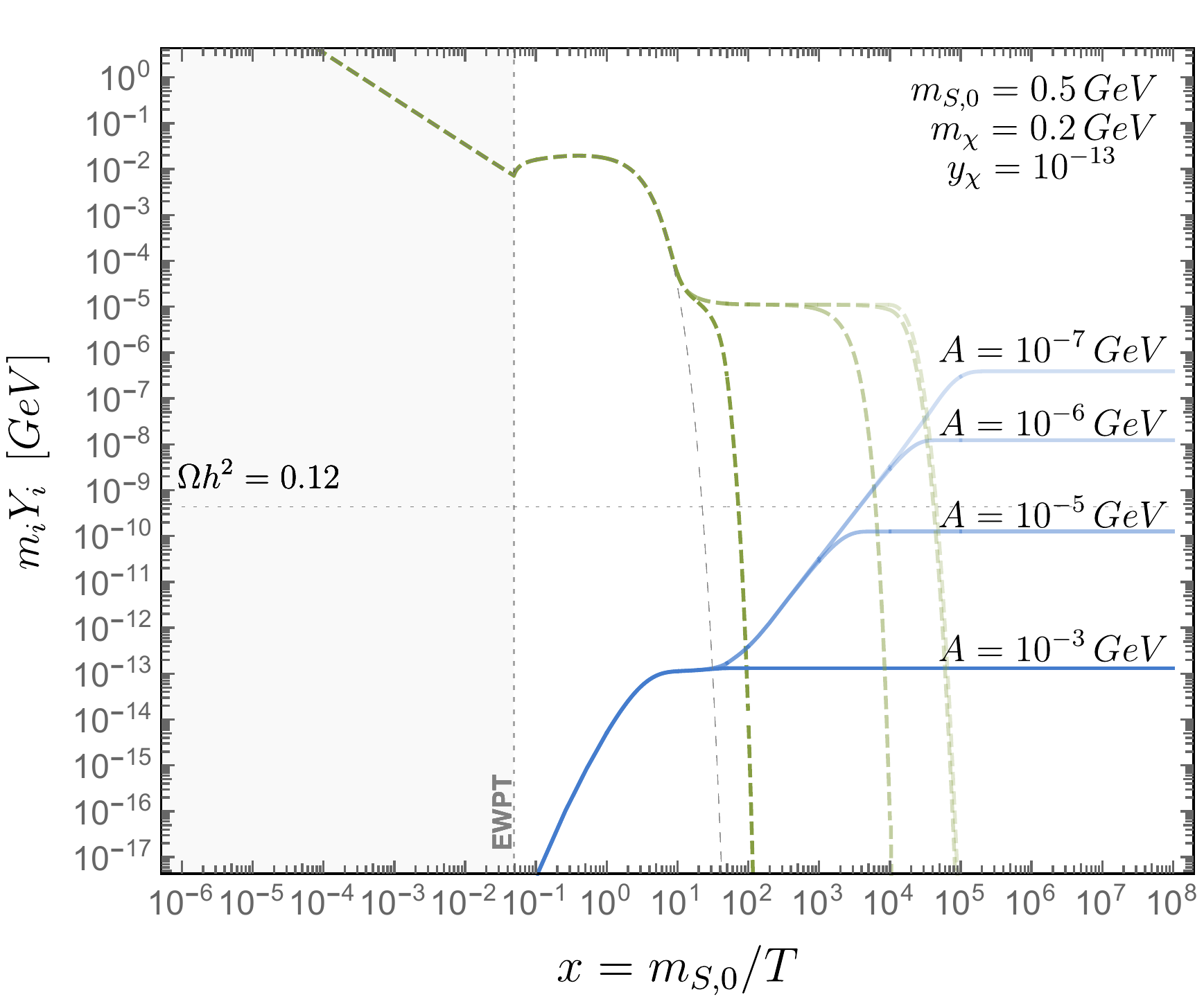}
		\caption{}
		\label{fig:Fig3a}
	\end{subfigure}
	%	\hfill
	\begin{subfigure}[b]{0.49\textwidth}
		\centering
		\includegraphics[width=1\textwidth]{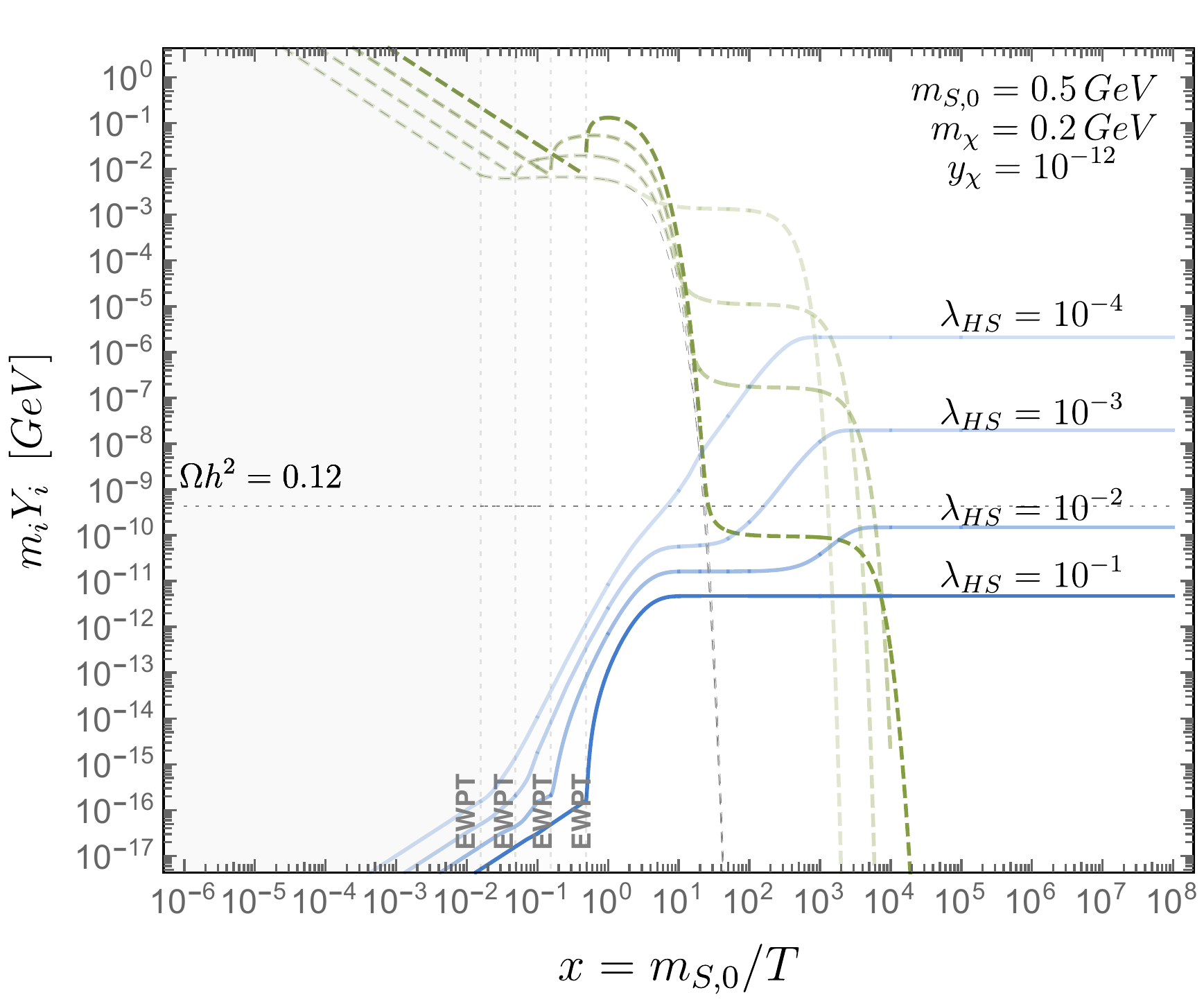}
		\caption{}
		\label{fig:Fig3b}
	\end{subfigure}
	\caption{Examples of yields evolution when the $\chi$
		production is dominated by the late time decay of $S$. {\bf
			(a)} Dependence on the trilinear coupling $A$, which (for
		fixed $\lambda_{HS}$) governs the branching ratio of $S$
		decay to $\chi$ and to SM particles. Here the freeze-out of
		$S$ proceeds as for usual WIMP, with decoupling at $x\sim
		20$. {\bf (b)} Dependence on the portal coupling
		$\lambda_{HS}$ (for fixed $A$). Lowering $\lambda_{HS}$
		leads to smaller mass, due to the EWSB contribution, and
		also earlier freeze-out with larger $Y_S$ which then
		translates to larger $\chi$ population. Note that in this
		plot the relation between $x$ and time/temperature is
		different for different lines.}
	\label{fig:Y3}
\end{figure}

%\begin{figure}[t]
%	\centering
%	\includegraphics[width=0.49\textwidth]{figs/Fig3a.pdf}
%	\includegraphics[width=0.49\textwidth]{figs/Fig3b.pdf}
%	\caption{Examples of yields evolution when the $\chi$ production is dominated by the late time decay of $S$. \textit{Left:} dependence on the trilinear coupling $A$, which (for fixed $\lambda_{HS}$) governs the branching ratio of $S$ decay to $\chi$ and to SM particles. Here the freeze-out of $S$ proceeds as for usual WIMP, with decoupling at $x\sim 20$. \textit{Right:} dependence on the portal coupling $\lambda_{HS}$ (for fixed $A$). Lowering $\lambda_{HS}$ leads to smaller mass, due to the EWSB contribution, and also earlier freeze-out with larger $Y_S$ which then translates to larger $\chi$ population. Note that in this plot the relation between $x$ and time/temperature is different for different lines.}  
%	\label{fig:Y3}
%\end{figure}

\subsubsection{Scan setup and results}

A numerical scan of the model parameter space has been conducted using
\texttt{MultiNest}~\cite{Feroz:2008xx} to direct the scan towards
values of the relic density within $2\sigma$ of the standard result
from the Planck Collaboration~\cite{Aghanim:2018eyx} $\Omega h^2 =
0.1198 \pm 0.0012$ that we set as an allowed range.\footnote{We used
	an additional $10\%$ theoretical uncertainty on our numerical
	results.} The private code \texttt{BayesFITS}, automatically created
using routines from
\texttt{SARAH}~\cite{staub_sarah_2008,Staub:2012pb,Staub:2013tta} is
used to interface it with the \texttt{Mathematica} implementing the
approach discussed above which we use to evaluate the relic
density. The details of the parameter ranges are given in
Table~\ref{tab:ranges}.

\begin{table}[h!]
	\begin{center}
		\begin{tabular}{|c|c|c|c|}
			\hline
			\rule{0pt}{1.25em}
			\textbf{Parameter} &  \textbf{Description} & \textbf{Range} &  \textbf{Prior}\\[0.15em]
			\hline
			\hline
			$\mu_\chi$~(GeV)${}$  & Dark matter Lagrangian mass & $\,0.005,\,50$ & Log \\
			$\mu_S$~(GeV$){}$ &Dark Higgs boson Lagrangian mass	 &$\,0.100,\,50$ & Log  \\
			$A$~(GeV) & Trilinear mixing	  &$\,10^{-8},\,10^{-2}$ & Log  \\
			\hline
			\rule{0pt}{1.5ex}
			$\lambda_{HS}$ & Quartic mixing	  &$\,10^{-8},\,10^{-2}$ & Log  \\
			$\ysx$ & Dark matter Yukawa &$\,10^{-14},\,10^{-8}$ & Log  \\
			$\lambda_{S}$ & Dark Higgs self-coupling &$\,10^{-4},\,1$ & Log  \\
			\hline
			
		\end{tabular}
		\caption{Ranges of the parameters of the model
                  analysed in this scan.  Dimensionful quantities are
                  given in GeV. }
		\label{tab:ranges}
	\end{center}
\end{table}%

\begin{figure}[h!]
	\centering
	\includegraphics[width=0.7\textwidth]{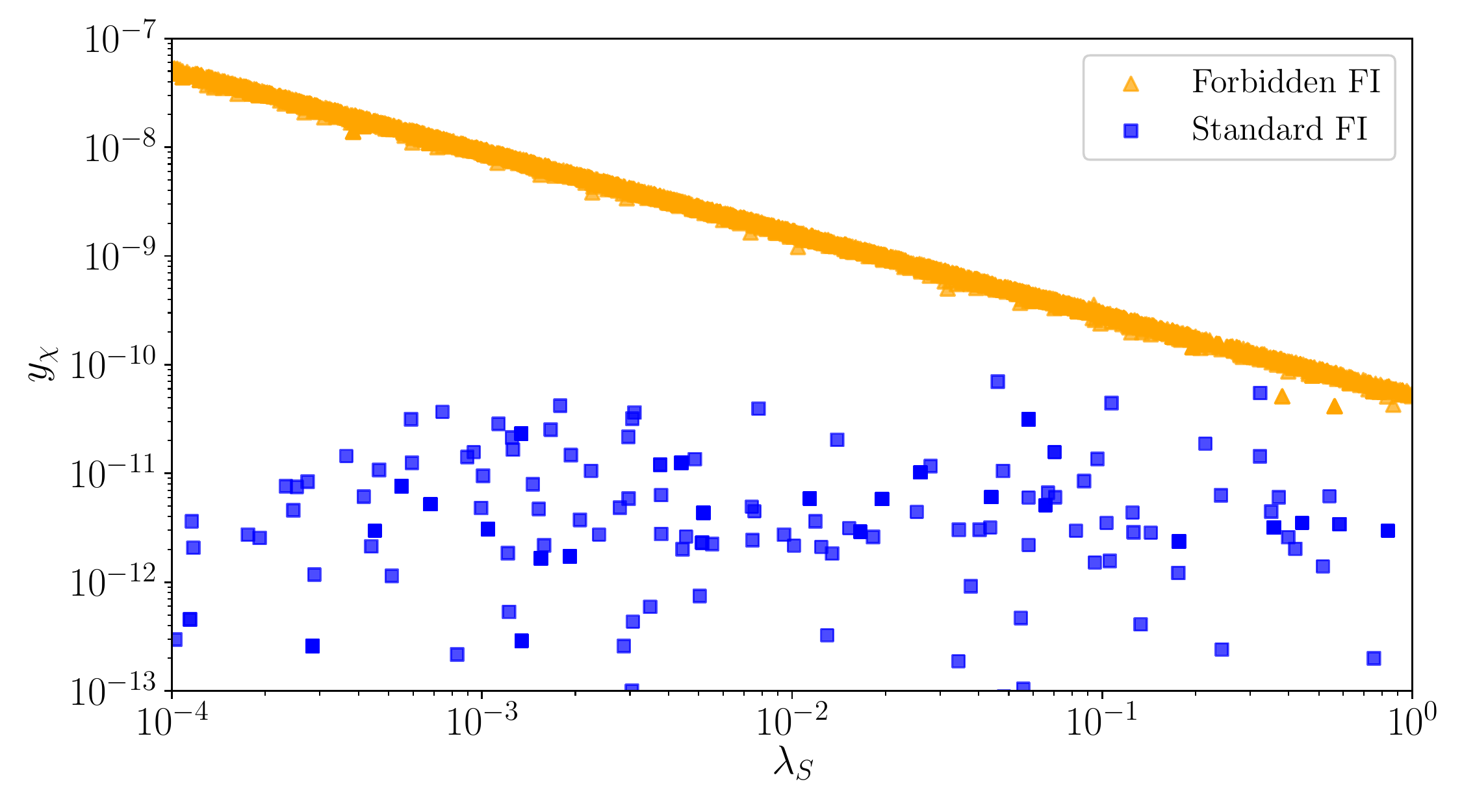}
	\caption{Points satisfying the observed relic density at 95\%CL in the plane $\lambda_S-\ysx$ for  $\mSV<2m_{\chi}$ (orange) and   $\mSV>2m_{\chi} $ (blue).} 
	\label{fig:OmScan}
\end{figure}

In Fig.~\ref{fig:OmScan} we show the points in the scan that satisfy
the DM relic density constraint. As before, blue colour indicates the
standard freeze-in regime and the beige one the forbidden regime. It
is apparent that these two regimes exhibit very distinct patterns. In
particular, as discussed in a previous section, the standard freeze-in
is in most cases not sensitive to the value of self-coupling
$\lambda_S$. It also requires very low values of the Yukawa coupling;
otherwise DM is overproduced. In contrast, the forbidden regime is
highly sensitive to $\lambda_S$, as expected. Indeed, the smaller the
self-coupling, and therefore the thermal mass, the earlier the
production stops and therefore the larger $\ysx$ is needed to obtain
the correct relic abundance of DM. Nevertheless, it is a new,
interesting regime that is generically present in our scans and
additionally leads to a freeze-in DM interacting more strongly than in
the usually studied scenarios. An important comment is that, while we
explicitly enforce the consistency condition~\eqref{eq:smalltri} for
all the scan-based plots, our choice of parameters implies that most
of our points with low dark Higgs boson mass exhibit also a small
quartic mixing $\lambda_{SH}$. This is a direct consequence of
\eqs{eq:masses}.

%%%%%%%%%%%%%%%%%%%%%%%%%%%%%%%%%%%%%%%%%%%%%%%%%%%%%%%
%%%%%%%%%%%%%%%%%%%%%%%%%%%%%%%%%%%%%%%%%%%%%%%%%%%%%%%%%%%

\subsection{Experimental limits}
\label{sec:explimits}

%%%%%%%%%%%%%%%%%%%%%%%%%%%%%%%%%%%%%%%%%%%%%%%%%%%%%%%
\begin{figure}[t]
	\centering
	\includegraphics[width=0.8\textwidth]{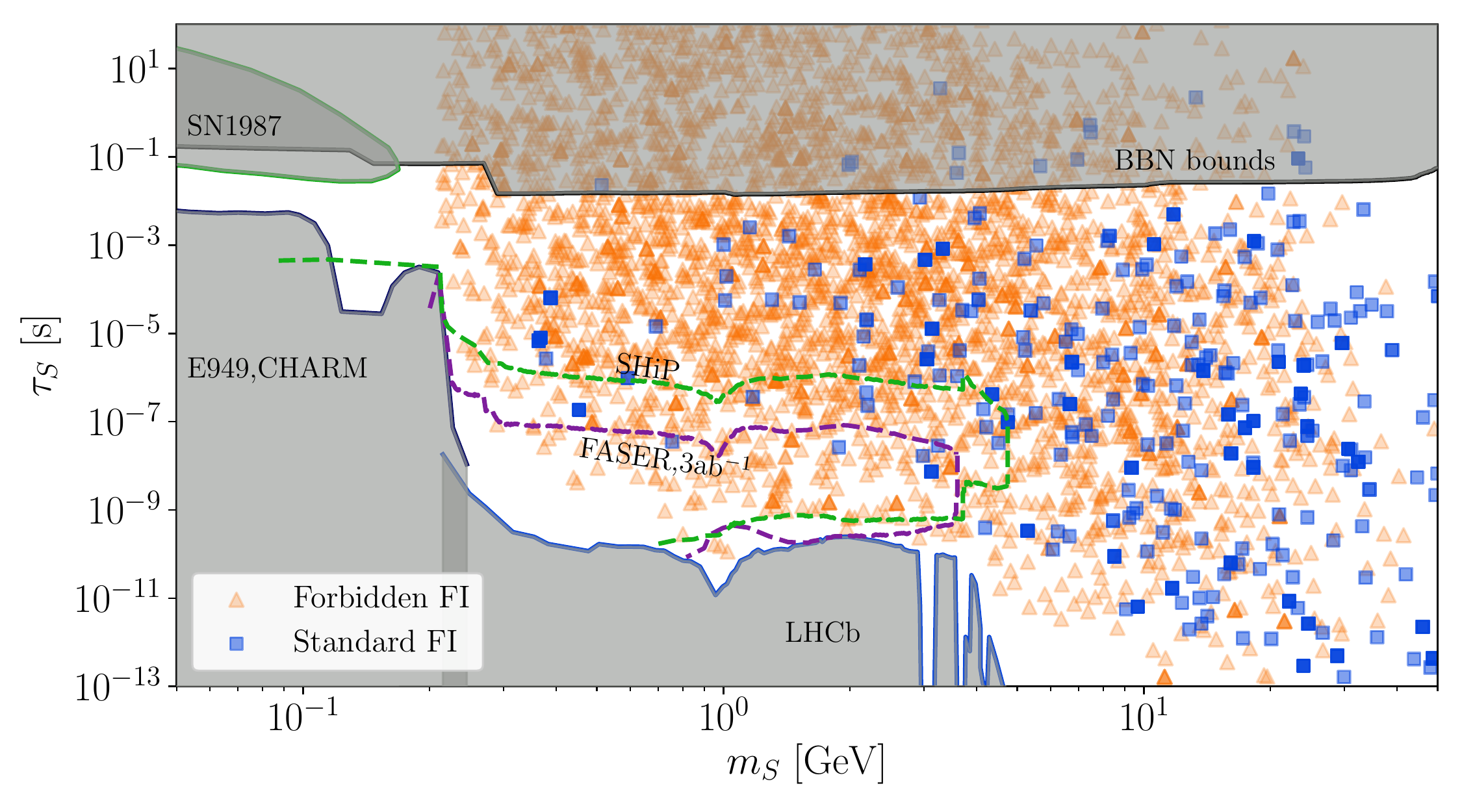}
	\caption{Experimental limits for our model, for points satisfying the observed relic density at 95\%CL in the plane $\mSV-\tau_S$ for  $\mSV<2m_{\chi}$ (orange) and   $\mSV>2m_{\chi} $ (blue).  } 
	\label{fig:Limits}
\end{figure}
%%%%%%%%%%%%%%%%%%%%%%%%%%%%%%%%%%%%%%%%%%%%%%%%%%%%%%%%%%%

%***\\
In dark Higgs models dark matter particles are largely out-of-reach of
current experiments due to their extremely small interactions with the
visible sector. The mixing of the scalars $h$ and $S$ induces,
however, interactions of $S$ with the SM particles which are
proportional to $\theta$, hence mediating the decay of $S$ to SM
particles (if kinematically allowed). Since $\theta$ is suppressed by
powers of $v_{S}/v$, the dark Higgs boson $S$ is typically
long-lived, as shown in \eqs{eq:lifetime} -- particularly for
low masses. In this case bounds from both colliders and fixed target
experiments~\cite{Beacham:2019nyx}, and for longer life-time, from
astrophysics~\cite{Fradette:2017sdd} apply. Such limits have
traditionally been very well-studied. We summarise them below and in
Figure~\ref{fig:Limits} which indicates the most relevant ones for our
setups. First of all, and apart from enforcing the proper dark matter
relic density, astrophysical bounds can be divided in two main
categories, and typically set an upper bound on the dark Higgs boson
lifetime, or equivalently a lower limit on its mixing angle with the
SM Higgs boson.

\begin{itemize}
	\item \textit{Cooling rate of the supernovae SN1987.} This limit uses the fact
	that the core of the nova is a thermal environment with temperature
	$T_{SN} \sim 30$ MeV where dark Higgs bosons can be produced and --
	if sufficiently feebly coupled -- escape the core and lead to a
	faster cooling of the supernova. Standard bounds for dark Higgs
	boson~\cite{Krnjaic:2015mbs} are derived from the requirement that
	the cooling rate from dark sector particles do not exceed the
	neutrinos
	one~\cite{Burrows:1986me,Burrows:1987zz,Raffelt:1996wa}.\footnote{While
		we use here the results from~\cite{Krnjaic:2015mbs}, this should be
		considered only an order of magnitude calculation. Note, however,
		that for the parameter space presented in Sec.~\ref{sec:Model},
		this bound is not directly relevant as can be seen in
		Figure~\ref{fig:Limits}.}
            \item \textit{Bounds from enforcing a successful big bang
                nucleosynthesis.} We use the recent bounds
              from~\cite{Fradette:2017sdd} which are derived from the
              same Lagrangian as in Sec.~\ref{sec:Model}. In the lower
              mass range (below the $\pi$-meson mass threshold) the
              dominant bounds are derived by constraining the entropy
              injections from the $e^+ e^-$/ $\mu^+ \mu^-$ decays of
              the dark Higgs boson. Once dark Higgs boson
              annihilation/decay into hadrons becomes accessible, more
              stringent bounds arise from preventing neutron-proton
              ratio to differ significantly from $1/6 \sim 1/7$ due to
              the $p \leftrightarrow n$ meson-mediated
              interaction. Finally, for heavy enough dark Higgs boson,
              direct baryon/anti-baryon production become the dominant
              decay channel of $S$. The subsequent anti-baryon
              annihilation with the ambient proton and neutron
              population further modifies the proton-neutron
              ratio. This limit dominates above the di b-quark
              threshold. An important comment is that this limits depends on the dark Higgs bosons
              abundance $Y_S$, however given our restriction
              \eqs{eq:smalltri}, dark Higgs bosons abundance typically
              freezes-out earlier than in~\cite{Fradette:2017sdd} which
              implies that the relativistic abundance is maintained
              for larger masses. Altogether, modifying $\lambda_{HS}$
              only changes the limits by an $\mathcal{O}(1)$ factor,
              as can be seen in~\cite{Fradette:2017sdd}. This is a
              simple consequence of the fact that, in order to avoid a
              significant modification of the $p/n$ ratio, one relies
              on ensuring that the dark Higgs boson decay
              \textit{before} BBN. The limit then roughly depends on
              the exponentially suppressed initial abundance $Y_S \exp
              (- t_{p/n}/\tau_S)$ where $t_{p/n} \sim 2.6 s $ is the
              freeze-out time of the proton/neutron ratio.\footnote{
                This behaviour is clearly illustrated in Figure 2,
                from ref.~\cite{Fradette:2017sdd}}
\end{itemize}
The second class of constraints arises from colliders and beam-dump experiments, and typically sets a lower bound on the dark Higgs boson life-time.
%\ld{Include here~\cite{Schmidt-Hoberg:2013hba,Clarke:2013aya} ? }
% Such decays are tightly constrained~\cite{Schmidt-Hoberg:2013hba,Clarke:2013aya}
%to $\theta \lesssim 10^{-4} - 10^{-2}$, for masses $\mSV = 0.1 \, - \, 10 \; \GeV$. This constraint is in agreement with our assumptions, as 
%$\theta$ is suppressed by powers of $v_{S}/v$ (or, equivalently, $A/v$).
%
\begin{itemize}
	\item \textit{ Limits from dark Higgs boson production and decay}. Based on the original ALP
	searches in CHARM~\cite{Bergsma:1985qz}, these limits have been recently updated with a	better modelling of the dark Higgs boson lifetime in the challenging
	region of $m_S$ around  $1$~GeV in~\cite{Winkler:2018qyg}. Note that
	we have included the projected limits from SHiP at $2\times 10^{20}$
	proton-on-target~\cite{Alekhin:2015byh} as a long-term prospect.
	Similarly, and as an example of limits from LHC-based experiments,
	we have included a projection for FASER phase 2 at the HL-LHC
	from~\cite{Feng:2017vli}. Notice that these next generation
	experiments have the potential to start probing the relevant
	parameter space.
	\item \textit{ Precision physics in meson decays.} In the lower mass
	range, the dominant limits arise from the meson decay $K^+
	\rightarrow \pi^+ \nu\nu$ studied in the E949
	experiment~\cite{Artamonov:2008qb}. Finally, for the heavier mass
	range -- corresponding to intermediate masses around $1$~GeV -- the
	main constraints come from searches for visible decay of B-meson by
	the LHCb collaboration~\cite{Aaij:2016qsm}. In both cases, we use
	the recasted bound from~\cite{Winkler:2018qyg}.
\end{itemize}
Note that in the long term, several planned experiment have the
potential to greatly improve the limits in this mass
range~\cite{Beacham:2019nyx}. LHC-based experiments, such as FASER,
MATHUSLA~\cite{Curtin:2018mvb} or CODEX-b~\cite{Gligorov:2017nwh} are
particularly interesting in that the decay of Higgs boson mediated
through the quartic mixing $\lambda_{HS}$ can significantly enhanced
the detection prospects as they are not tied to the mixing angle
per-se but only to $\lambda_{SH}$ (hence the invisible branching ratio
of the SM Higgs). Saturating the limits from invisible Higgs decay
then leads to orders of magnitude improvements, particularly in the
case of MATHUSLA or CODEX-b~\cite{Beacham:2019nyx}.

\section{Conclusion}
In this article we studied the forbidden freeze-in
regime. Building on a standard decay-mediated freeze-in scenario, we
focused on the case where the decaying mediator field couples strongly
enough to the SM thermal bath to develop a significant thermal mass at
high temperature. This strongly modifies existing predictions, and in
particular leads to a particularly interesting regime of forbidden
freeze-in, where the decay into DM particles is kinematically
forbidden in the vacuum but is allowed to proceed in the thermal bath.
%into account the thermal mass of bath particle(s) that decays to DM ones, even when said decays are . 
% \bl{D: Not particularly happy with this...a few more things needed!.}

In Sec.~\ref{sec:forbfreeze-in}, we described in some detail
the effect of including a sizeable thermal mass of the
mediator. Assuming that the main production channel of DM is the
decays of a bath particle into a pair of DM particles, we showed that
freeze-in can be dominant at  both high and low temperatures,
depending on the dimension of the operators that couple the DM to the
bath particle. Although the $d>4$ operators show high-temperature
dominance of DM production, this is different from  the standard
freeze-in case at high temperatures since the dominance does not happen due to
the kinematics of the production process, but due to the thermal mass
of the bath particle. Comparing the forbidden with the standard case
of high-temperature freeze-in, we showed that the forbidden freeze-in is
generally less efficient, leading to a stronger coupling between the DM
particle and the mediator.
For the case of operators with $d\leq4$  we showed that the production is dominant
at lower temperatures  close to the DM mass. In this case the
scale of DM production is insensitive to the scale of inflation and
reheating, similarly to the case of standard
``freeze-out''. Furthermore, the relic abundance is ultimately almost
insensitive to the DM mass and the coupling responsible for the DM
production can take significantly larger values than in the standard
freeze-in scenario.

As a concrete example we studied a scalar portal model where the DM
(assumed to be a Dirac fermion) is coupled only to a scalar which in
turn is coupled to the SM Higgs boson field.  In Sec.~\ref{sec:DarkHiggs} 
we showed the effect that the scalar thermal mass has on the production of DM. We 
studied in detail the solution of the coupled Boltzmann equations for the DM particle
and the mediator and discussed various possible types of the evolution of DM relic 
density.  We also performed a scan of the parameter space of the model at hand and 
presented the region where the observed relic abundance can be obtained. Focusing on 
the same model, in Sec.~\ref{sec:DarkHiggs} we discussed its experimental search
prospects. Since the DM coupling to the SM particles is expected to be
extremely suppressed (due to the small Yukawa coupling and the small
mixing angle between the portal and Higgs boson fields) this model can
be mostly probed by searching for a long-lived scalar mediator.  We
showed the impact of all the relevant bounds on the parameter space,
including BBN, LHCb, CHARM, as well as astrophysical bounds for the
presence of a light scalar field coupled to the Higgs boson. Also, we
discussed the reach of upcoming fixed-target experiments (SHiP and
FASER) and showed what part of the parameter space they will be able
to probe.

As we have already pointed-out, the forbidden freeze-in regime 
is a general feature of the freeze-in mechanism. It greatly expands the
parameter space in models where otherwise the DM cannot be produced by the decays 
of bath particle. Therefore, the analysis performed in this work not only
provides new interesting viable regions of the Higgs portal model
but may also bring some insight into how the forbidden freeze-in
works in general. Our results also strongly suggest that it would be
interesting to re-examine the dark matter abundance in other types of
freeze-in models in order to uncover their respective forbidden freeze-in regimes.

\bigskip
%%%%%%%%%%%%%%%%%%%%%%%%%%%%%%%%%%%%%%%%%%%%%%%%%%%%%%%%%%%%%%%%%%%%%%%%%%%%%%%%
\noindent \textbf{Acknowledgments}
\medskip

LD, DK and LR are supported in part by the National Science Centre, Poland, research grant No. 2015/18/A/ST2/00748.
AH is supported in part by the National Science Centre, Poland, research
grant No. 2018/31/D/ST2/00813. LR is also supported by the project
“AstroCeNT: Particle Astrophysics Science and Technology Centre”
carried out within the International Research Agendas programme of the
Foundation for Polish Science financed by the European Union under the
European Regional Development Fund.

%%%%%%%%%%%%%%%%%%%%%%%%%%%%%%%%%

%%%%%%%%%%%%%%%%%%%%%%%%%%%%%%%
\newpage 
\bibliography{refs}{}
\bibliographystyle{JHEP}                        

\end{document}